# A Rule-Based Change Impact Analysis Approach in Software Architecture for Requirements Changes

ARDA GOKNIL[1], IVAN KURTEV[2], KLAAS VAN DEN BERG[3]

[1]SnT Centre, University of Luxembourg, Luxembourg
[2]Altran, Limburglaan 24, 5652 AA, Eindhoven, the Netherlands
[3]University of Twente, 7500 AE Enschede, the Netherlands
[1]arda.goknil@uni.lu, [2]ivan.kurtev@altran.com, [3]vdberg.nl@gmail.com

**Abstract:** Software systems usually operate in a dynamic context where their requirements change continuously and new requirements emerge frequently. A single requirement hardly exists in isolation: it is related to other requirements and to the software development artifacts that implement it. When a requirements change is introduced, the requirements engineer may have to manually analyze all requirements and architectural elements for a single change. This may result in neglecting the actual impact of a change. We aim at improving change impact analysis in software architecture for requirements changes by using formal semantics of requirements relations, requirements changes and traces between Requirements & Architecture. In our previous work we presented a technique for change impact analysis in requirements. The technique uses the formal semantics of requirements relations and changes. Its output is a set of candidate requirements for the impact with proposed changes and a propagation path in the requirements model. In this paper we present a complementary technique which propagates requirements changes to software architecture and finds out which architectural elements are impacted by these changes. The formalization of requirements relations, changes and traces between R&A is used to determine candidate architectural elements for the impact of requirements changes in the architecture. The tool support is an extension of our Tool for Requirements Inferencing and Consistency Checking (TRIC). Our approach helps in the elimination of some false positive impacts in change propagation. We illustrate our approach in an industrial example which shows that the formal semantics of requirements relations, changes and traces enables the identification of candidate architectural elements with the reduction of some false positive impacts.

## 1 Introduction

Today's software systems usually operate in a dynamic business context where business goals often change. As a result, the requirements of software systems change continuously and new requirements emerge frequently. The integration of the new requirements and the adaptations to the deployed software systems are costly and time consuming. A single requirement hardly exists in isolation: it is related to other requirements and to the software development artifacts that implement it. Thus, even a simple change in a single requirement may have a significant effect on the whole system. Determining such an effect is usually referred to as *change impact analysis*. In this paper we focus on change impact analysis in software architecture for requirements changes.

The need for change impact analysis is observed in both requirements and software architecture. When a change is introduced to a requirement, the requirements engineer needs to find out if any other requirement related to the changed requirement is impacted. After determining the impacted requirements, the software architect needs to identify the impacted architectural elements by tracing the changed requirements to software architecture. It is hard, expensive and error prone to manually trace impacted requirements and architectural elements from the changed requirements. Several commercial tools such as IBM RequisitePro and DOORS use semi-structured format of requirements documents and traces between Requirements (R) and Architecture (A) with support for automatic change impact analysis. Although these tools capture requirements relations and traces between R&A (and also other artifacts) explicitly, the meaning of these relations and traces is often too general and the analysis results may be deficient [10]. For instance, when a requirement is changed in RequisitePro, all reachable elements are considered potentially impacted. The relation types (*tracedTo* and *tracedFrom*) provided by RequisitePro indicate only the direction of the relations and traces without their actual meaning. By using only the transitive closure of relations and traces, the software architect may conclude that all architectural elements are impacted. Without any additional semantic information about the requirements relations, changes, and traces, change impact analysis may produce a high number of false positive impacts. This situation is explained by Bohner [1] [2] [3] as *explosion of impacts without semantics*. It is concluded that change impact analysis must employ additional semantic information to increase the accuracy by eliminating false positives [2].

In our previous work [24] [22], we use a representation of requirements and relations among them as models conforming to a requirements metamodel. The metamodel contains concepts commonly found in the literature and that reflect how most requirements documents are structured. The main elements in the metamodel are requirements relations and their types. The semantics of these elements is given in First Order Logic (FOL) and allows two activities. First, new relations among requirements can be inferred from the initial set of relations. Second, requirements models can be automatically checked for consistency of the relations. The tool for *Requirements Inferencing and Consistency Checking* (TRIC) [24] [55] is developed to support both activities. The details about the metamodel, the semantics and the tool are already reported in our previous paper [24]. As a continuum, we presented another approach [21] that provides trace establishment between R&A by using architecture verification together with the semantics of requirements relations and traces. We use a trace metamodel with commonly used trace types. The semantics of traces is formalized in FOL. Software architectures are expressed in the Architecture Analysis and Design Language (AADL) [48]. AADL is provided with a formal semantics expressed in Maude [43] [42] which allows simulation and verification of architectures. We use semantics of requirements relations and traces to both generate/validate traces and generate/validate requirements relations. The approach is supported with a tool.

We extended our results with a technique [20] for change impact analysis in requirements models. The technique uses the formal semantics of requirements relations and a classification of requirements changes. Three activities for impact analysis in requirements models are supported. First, the requirements engineer proposes changes according to the change classification before implementing the actual changes. Second, the requirements engineer identifies the propagation of the changes to related requirements. Possible alternatives in the propagation are determined based on the semantics of change types and requirements relations. Third, possible contradicting changes are identified. TRIC is extended to support these activities. The tool automatically determines change propagation paths, checks consistency of the changes, and suggests alternatives for implementing the changes. There are different rationales for requirements changes (e.g., *refactoring* and *domain changes*). Our focus is the requirements changes fostered by the evolution of the business needs since they have an impact on other software development artifacts such as software architecture, detailed design and source code. We name these changes as *domain changes* [20].

The output of the change impact analysis approach in requirements models is a set of impacted requirements with proposed changes and a propagation path in the requirements model. The next step is to find out which architectural elements are impacted by these requirements changes. The software architect needs to decide which impacted requirement(s) in the propagation path should be traced to architecture to determine impacted architectural elements. In this paper we present a rule-based change impact analysis approach for software architecture using the formal semantics of requirements relations, changes, and traces between R&A. By having the formal semantics, we derive change impact rules to identify which parts of the architecture are candidate to be impacted by a requirements change. Our approach is automatic. It traverses the change propagation path in the requirements model to determine which impacted requirement(s) is traced to architecture. It helps in the elimination of some false positive impacts. We have extended TRIC for the tool support. We use AADL to model the architecture but our approach is independent of any architecture modeling approach. The architecture can be expressed with any other ADL or more generic notations such as UML. We only require traces between R&A.

This paper is structured as follows. Section 2 introduces an illustrative industrial example used in the following sections. In Section 3, we illustrate an example impact analysis with and without using semantics. Section 4 briefly introduces the elements of the requirements and trace metamodels. Section 5 presents the change classification in requirements. In Section 6, we summarize our change impact analysis approach in requirements. Although Sections 4, 5 and 6 are already given in our previous work, they are needed for understanding our approach. Section 7 explains our rule-based change impact analysis approach in architecture. The tool support is given in Section 8. In Section 9, we discuss our approach. Section 10 presents the related work and we conclude the paper in Section 11.

## 2   Running Example: Remote Patient Monitoring System

The approach will be illustrated with the Remote Patient Monitoring (RPM) system example. The RPM system was developed by a Dutch company and had already been implemented and running when we started studying the system. It has the following stakeholders: *patients*, *doctors*, and *a system administrator*. The main goal is to enable monitoring the patients' conditions such as blood pressure, heart rate, and temperature. For instance, a patient carries a sensor for measuring the body temperature and the values are transmitted to a central storage system. Table 1 gives some of the requirements for the system.

**Table 1** Some of the Requirements for the RPM System

| |
|---|
| **R1** The system shall measure temperature from a patient. |
| **R2** The system shall measure blood pressure from a patient. |
| **R3** The system shall measure blood pressure and temperature from a patient. |
| **R4** The system shall store patient temperature measured by the sensor in the central storage. |
| **R5** The system shall store patient blood pressure measured by the sensor in the central storage. |
| **R6** The system shall store data measured by sensors in the central storage. |
| **R7** The system shall warn the doctor when the temperature threshold is violated. |
| **R8** The system shall generate an alarm if the temperature threshold is violated. |
| **R9** The system shall show the doctor the temperature alarm at the doctors' computers. |
| **R10** The system shall store all generated temperature alarms in a central storage. |
| **R11** The system shall enable the doctor to set the temperature threshold for a patient. |
| **R12** The system shall enable the doctor to retrieve all stored temperature measurements for a patient. |
| **R13** The system shall enable the doctor to retrieve all stored temperature alarms for a patient. |
| **R14** The system shall store patient temperature measured by the sensor in the central storage and it shall warn the doctor when the temperature threshold is violated. |
| **R15** The system shall store patient Central Venous Pressure (CV Pressure) measured by the sensor in the central storage. |

In the paper we use some requirements changes for the RPM system. These changes are derived from the experience of the software engineer involved in the RPM system development. The RPM architecture is constructed by reverse engineering the source code. Figure 1 gives the overview of the RPM architecture in AADL. The explanation of the component abbreviations is given in Appendix 1. The reader is referred to Appendix 2 for the AADL graphical notation.

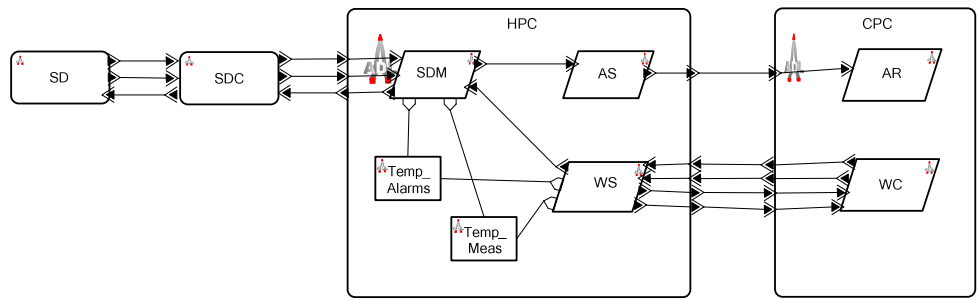

**Figure 1** Overview of the RPM Architecture

Figure 1 shows the most abstract components (*system* and *process* in AADL). These components contain other components not represented in Figure 1. The *SD*

*(Sensor Device)* component contains the patient sensors which perform measurements at a regular interval. The *SD* sends the measurements to the *HPC (Host Personal Computer)* component through the *SDC (Sensor Device Coordinator)*. The *SDC* is the ZigBee network coordinator. The details of the coordinating tasks are omitted. The *HPC* consists of the *SDM (Sensor Device Manager)*, *AS (Alarm Service)* and *WS (Web Server)* process subcomponents. The *SDM* stores the measurements and generated alarms in the data stores (*Temp_Alarms* and *Temp_Meas* for temperature alarms and measurements). The *WS* serves as a web-interface for the doctors. The *AS* forwards the alarms to the *CPC (Client Personal Computer)* component. The *CPC* is used by the doctors to monitor patients. The *AR (Alarm Receiver)* process in the *CPC* receives the alarms from the *AS* and notifies the doctor. The *WC (Web Client)* process uses the *WS* to retrieve the measurements and alarms stored by the *SDM*.

Figure 1 shows only systems and processes in the RPM architecture. AADL provides support for *thread* and *subprogram* components. The computation of the system is modeled as a subprogram and thread behavior. The RPM architecture has behavioral annexes for dynamic behavior of threads in each system component. For brevity, we do not include the behavioral annexes in this section.

## 3  The Importance of Semantics for Change Impact Analysis

The main challenge in change impact analysis is to find exactly the elements that need a change and to minimize the number of false positives. Figure 2 gives part of the RPM requirements and architecture with some trace links modeled in IBM RequisitePro. There are 'traced to' relations between requirements and 'traced from' relations connecting architectural components to the requirements they implement. *R3* gives two system properties (Measuring patient's blood pressure and Measuring patient's temperature). Assume a change request for introducing a new constraint "applying the so called oscillometric method" to the property "Measuring patient's blood pressure" in *R3*. After *R3* is updated by adding the new constraint, the property becomes the following: "The system should measure the blood pressure from the patient by applying the so called oscillometric method". By following direct and indirect relations without their meaning, it is found that all requirements (*R1*, *R2*, *R3*, *R4*, *R5* and *R6*) and architectural elements (*C1*, *C2*, *C3*, *C4*, *C5* and *C6*) are suspects for the impact.

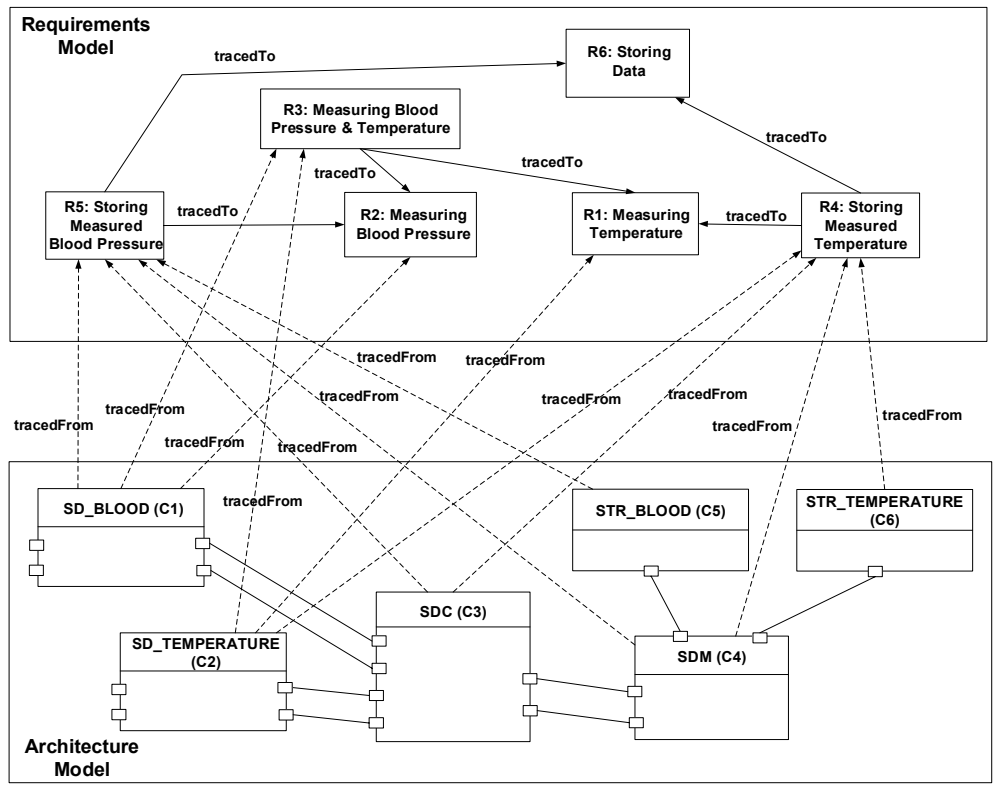

**Figure 2** Part of the RPM Requirements and Architecture with RequisitePro

By having the semantics of the change, requirements relations and traces, some of the suspects are identified as false positives. The requested change is adding a new constraint to the property (measuring blood pressure) in *R3* which does not have any impact on the directly related *R1* since *R1* gives another property (measuring temperature) of *R3*. On the other hand, *R2* is impacted since it gives exactly the same property to which the constraint is introduced in *R3* (*R3* is a collection of the properties given in *R1* and *R2*). *R5* gives a property (storing measured blood pressure) which requires the impacted property (measuring blood pressure) in *R2* but storing the blood pressure does not consider how to measure the blood pressure. Therefore, *R5* is not impacted by the change in *R2*. The set of impacted elements is reduced at an early stage and the change is not propagated to *R1*, *R4*, *R5* and *R6*. Only the *Sensor Device for Blood Pressure (SD_BLOOD)* implements the impacted property "Measuring patient's blood pressure" in *R3* and *R2* while *SD_TEMPERATURE* implements a different property "Measuring Temperature" in *R3*. Therefore, *SD_BLOOD* is the only candidate for the impact in the architecture. If the (semi-)automatic change impact analysis does not utilize semantic information as described above, the software architect has to make the reasoning by himself.

## 4  Modeling Requirements and Traces

In this section we give a brief description of our requirements and trace metamodels. Figure 3 shows the trace metamodel together with parts of the requirements and architecture metamodels. We use AADL to specify architectural models. A fragment of the AADL metamodel is given at the bottom of Figure 3.

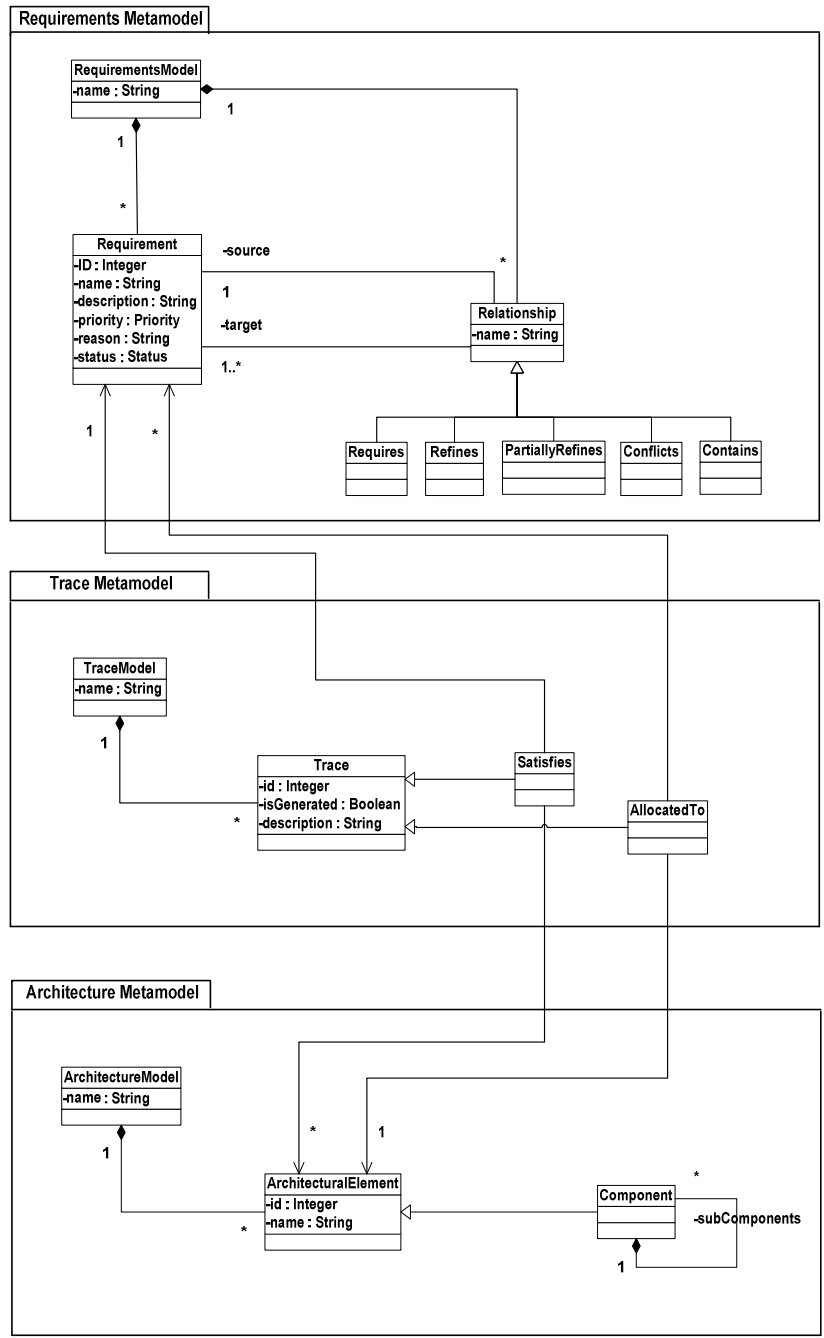

**Figure 3** Requirements, Trace and Architecture Metamodels [21]

Our requirements metamodel contains common entities identified in the literature for requirements models. In order to construct our requirements metamodel we investigated and benefited from several approaches which are commonly used to define and represent requirements: goal-oriented [54] [40], aspect-driven [47], variability management [38], use-case [9], domain-specific [44] [33], and reuse-driven techniques [36]. The main elements in the requirements metamodel are *Requirement* and *Relationship* (see Figure 3). The metamodel defines the *Requirement* entity with its attributes and relations between requirements. Based on [49] we define a requirement as follows:

- **Definition 1.** *Requirement*: A requirement is a description of a system property or properties which need to be fulfilled.

We identified five types of relations: *requires*, *refines*, *partially refines*, *contains*, and *conflicts*. In the literature, these relations are informally defined as follows.

- **Definition 2.** *Requires relation*: A requirement $R_1$ *requires* a requirement $R_2$ if $R_1$ is fulfilled only when $R_2$ is fulfilled.
- **Definition 3.** *Refines relation*: A requirement $R_1$ *refines* a requirement $R_2$ if $R_1$ is derived from $R_2$ by adding more details to its properties.
- **Definition 4.** *Contains relation*: A requirement $R_1$ *contains* requirements $R_2$ ... $R_n$ if $R_2$ ... $R_n$ are parts of the whole $R_1$ (part-whole hierarchy).
- **Definition 5.** *Partially refines relation:* A requirement $R_1$ *partially refines* a requirement $R_2$ if $R_1$ is derived from $R_2$ by adding more details to properties of $R_2$ and excluding the unrefined properties of $R_2$.
- **Definition 6.** *Conflicts relation*: A requirement $R_1$ *conflicts with* a requirement $R_2$ if the fulfillment of $R_1$ excludes the fulfillment of $R_2$ and vice versa.

We modeled the RPM textual requirements and their relations in TRIC according to the semantics of the relation types. Figure 4 gives an instance of the requirements metamodel for R5 and R15 (R15 refines R5).

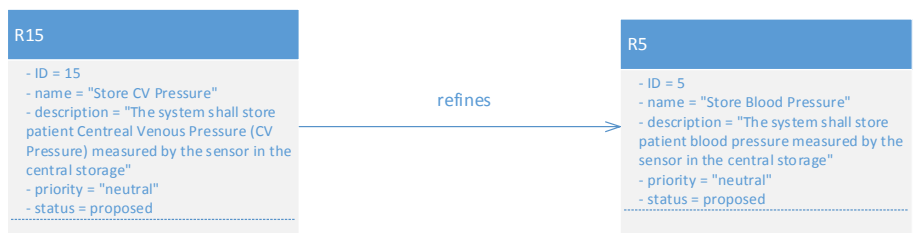

**Figure 4** Instance of the Requirements Model for R5 and R15

**R5** *The system shall store patient blood pressure measured by the sensor in the central storage.*

**R15** *The system shall store patient Central Venous Pressure (CV Pressure) measured by the sensor in the central storage.*

Figure 5 shows a part of the RPM requirements model in our approach.

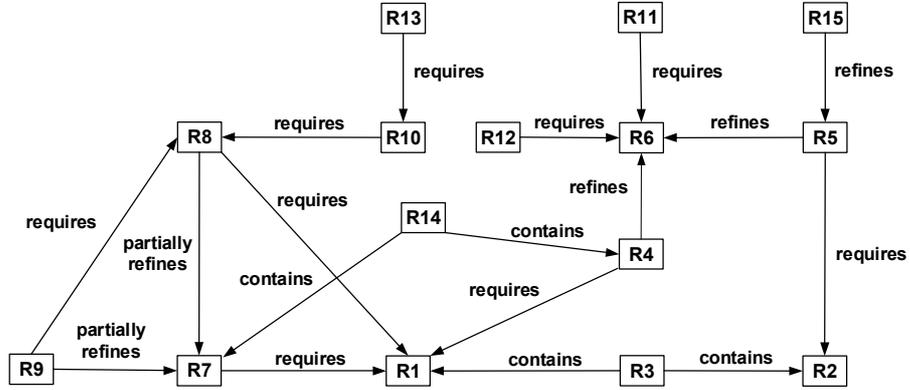

**Figure 5** Part of the Requirements Model for the RPM System in TRIC

The solid arrows indicate the relations given by the requirements engineer. For simplicity, we did not include the requirements properties and the relations inferred by TRIC.

Traces are collected in models that conform to the trace metamodel which contains two trace types *Satisfies* and *AllocatedTo*. The *AllocatedTo* and *Satisfies* relations are defined as follows [44] [46] [56]:

**Definition 7.** *AllocatedTo trace*: A requirement *R* is *allocated to* a set of architectural elements *E* if the system properties related to *E* are supposed to fulfill the system properties given in *R*.

**Definition 8.** *Satisfies trace*: A set of architectural elements *E satisfies* a requirement *R* if the system properties related to *E* fulfill the system properties given in *R*.

The definitions of the types of traces and requirements relations given above are informal. The semantics of the requirements relations and traces is formalized in first-order logic (FOL). Supplementary Material A[1] presents the formal semantics of requirements and relations. For the detailed description of the formal semantics of traces, the reader is referred to our previous work [21]. In this paper we will use the informal definitions given above.

---

[1] http://people.svv.lu/goknil/supplementary/SupplementaryMaterialA.pdf

## 5 Classification of Changes in Requirements

Our approach uses a classification of requirements changes. Requirements changes are analyzed and classified based on an assumption about a very general structure of a textual requirement. The change types are formalized by giving their effects in terms of changes in the formula that represents a requirement. The complete formal semantics of the change types and their rationale can be found in Supplementary Material B[2] and in our previous paper [20].

### 5.1 Structure of a Textual Requirement

We need to consider a general enough structure of a requirement to determine the granularity of changes that can be applied. Our definition of a requirement is "a textual requirement is a description of a property or properties which must be exhibited by the system" (see Figure 6).

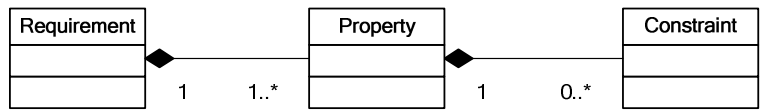

**Figure 6** Structure of a Textual Requirement based on the Definition of a Requirement

**Example:** *Structure of Requirement based on the definition*

**R2** *The system shall measure blood pressure from a patient.*

We can identify the following structure of R2 by following Figure 6:

**Property:** The system shall provide the functionality of measuring data from a patient.

**Constraint:** The measured data is patient's blood pressure.

### 5.2 Change Types for Requirements Models

The change types for requirements models are derived from the structure in Figure 6 and from the requirements metamodel in Figure 3 (see Table 2).

**Table 2** Requirements Change Types

| Change Types |
|---|
| ▪ Add a New Requirements Relation |
| ▪ Delete Requirements Relation |
| ▪ Update Requirements Relation |
| ▪ Add a New Requirement |
| ▪ Delete Requirement |
| ▪ Update Requirement |
|     o Add Property to Requirement |

---

[2] http://people.svv.lu/goknil/supplementary/SupplementaryMaterialB.pdf

|   |   |
|---|---|
| o | Add Constraint to Property of Requirement |
| o | Change Property of Requirement |
| o | Change Constraint of Property of Requirement |
| o | Delete Property of Requirement |
| o | Delete Constraint of Property of Requirement |

The change types in Table 2 do not address why a change needs to be performed in the requirements model, that is, what is the rationale of changes. The change rationale is important. It is a factor in identifying impacted architectural elements. In our approach we take the following definitions for the types of change rationale:

***Refactoring.*** Refactoring is a change (changes) to improve the structure of the requirements model without modifying the overall system properties [15]. These changes do not affect the properties in the whole model.

***Domain Changes.*** Domain changes are the changes in order to modify the overall system properties in the requirements model. These changes do affect the properties in the whole requirements model.

### 5.3  Semantics of Requirements Changes

In this section we sketch the formalization of requirements with the semantics of the change "Add Constraint to Property of Requirement".

*Formalization of Requirements*

We assume the general notion of requirement being "a property which must be exhibited by a system". We express the property as a formula P in FOL. We assume that requirements can always be expressed in the universal fragment of FOL and a requirement is expressed as a formula $\forall x \varphi$ with $\varphi$ in conjunctive normal form (CNF). If the formula $\varphi$ is a closed formula, then the universal quantifiers can be dropped. It is also possible that the formula contains free variables.

According to the model theoretic semantics of FOL, the truth value of P is determined in a *model* $\mathcal{M}$ by using an interpretation for the function and predicate symbols in P.

Let $\mathcal{F}$ be a set of function symbols and $\mathcal{P}$ a set of predicate symbols, each symbol with a fixed arity. A model $\mathcal{M}$ of the pair ($\mathcal{F}, \mathcal{P}$) consists of the following items [29]:

- a non-empty set $A$, the *universe of concrete values*

- for each $f \in \mathcal{F}$ with n arguments, a function $f^{\mathcal{M}} : A^n \to A$
- for each $P \in \mathcal{P}$ with n arguments, a set $P^{\mathcal{M}} \subseteq A^n$.

A satisfaction relation between the model $\mathcal{M}$ and the formula P holds:

(1)     $\mathcal{M} \models_{\ell} P$

if P evaluates to True in the model $\mathcal{M}$ with respect to the environment $\ell$ (i.e., a look-up table for free variables in P). The model $\mathcal{M}$ together with $\ell$ in which P is true represents a system $s$ that satisfies the requirement. From now on, all the formulae P that express properties will be in the form where ($\forall x = \forall x_1 \forall x_2 \ldots \forall x_k$):

(2)     $P = \forall x \, (p_1 \wedge \ldots \wedge p_n)$, where $n \geq 1$

$p_n$ is a disjunction of literals which are atomic formulae (atoms) or their negation. An atomic formula is a predicate symbol applied over terms. In the rest of the paper we use the notation $(p_1 \ldots p_n)$ for $(p_1 \wedge \ldots \wedge p_n)$.

In the following we give the semantics of the change "Add Constraint to Property of Requirement".

*Add Constraint to Property of Requirement*

Let R be the requirement before adding the constraint $ct$ to the property $pt$, and $R^l$ be the requirement after adding the constraint $ct$ to the property $pt$. P and $P^l$ are formulas for R and $R^l$. P is in conjunctive normal form as follows:

(3)     $P = \forall x \, ((p_1 \ldots p_n) \wedge (q_1 \ldots q_m))$;   $m, n \geq 1$

Let $p_1^l, p_2^l, \ldots, p_{n-1}^l, p_n^l$ be disjunction of literals such that $\forall x \, (p_j^l \to p_j)$ for all $j \in 1..n$

R becomes $R^l$ after adding $ct$ to the property $pt$ of R iff $P^l$ is derived from P by replacing every $p_j$ in P with $p_j^l$ for $j \in 1..n$ such that the following two statements hold:

(4)     $P^l = \forall x \, ((p_1^l \ldots p_n^l) \wedge (q_1 \ldots q_m))$;  $n \geq 1, m \geq 0$

(5)     $(\neg (\forall x \, (p_j \to p_j^l)))$ is satisfiable for all $j \in 1..n$

For the formulas $\forall x \, (p_1^l \ldots p_n^l)$ and $\forall x \, (q_1 \ldots q_m)$, if any variable universally quantified in one of the formulas appears free in the second formula, the free variable is renamed. If any variable in $\forall x\,(q_1 \ldots q_m)$ appears in $\forall x\,(p_1^1 \ldots p_n^1)$ with a different valuation, the variable in $\forall x\,(p_1^1 \ldots p_n^1)$ is renamed.

This change is similar to refining a requirement (see the *refines* relation in Supplementary Material A). The idea behind the change is to make the requirement more restrictive by adding a constraint. Therefore, the requirement after the change is a refinement of the requirement before the change.

## 6 Change Propagation in Requirements Models

Our approach in this paper heavily depends on the output of our previous work [20] for change impact analysis in requirements. *Change propagation in requirements* lies at the heart of our previous work. It is a process of deducing new proposed changes in a requirements model based on an initial set of proposed changes. The requirements relations and change types are used to determine if a change in a requirement has an impact (is propagated) on directly related requirements. In this section we give a brief overview of our previous work [20] for change propagation in requirements. The reader is referred to [20] for the details such as how we avoid cyclic change impacts in change propagation and how multiple parallel requirements changes are taken into account.

The requirements engineer proposes a change for a requirement based on the change classification. Change alternatives are automatically provided for each directly related requirement by our tool TRIC. The requirements engineer propagates the change from the impacted requirement to the directly related requirements by choosing changes among the provided alternatives. 'No Impact' is automatically identified for some of the related requirements. Change propagation is done step-by-step. Once a change is propagated to directly related requirements, the next step is to propagate the change to indirectly related requirements from directly related impacted requirements.

Change alternatives are determined based on the semantics of the change types, change rationale and requirements relations. Table 3 gives change alternatives for some of the change types (see the PhD thesis [18] for the complete table). We consider the changes with the *domain change* rationale since they have an impact on other software development artifacts such as software architecture and source code.

**Table 3** Change Impact Alternatives for Some Change Types with the Domain Change Rationale

| Change Types | Requirements Relation Types | | | | |
|---|---|---|---|---|---|
| | $R_i$ contains $R_k$ | $R_i$ refines $R_k$ | $R_i$ partially refines $R_k$ | $R_i$ requires $R_k$ | $R_i$ conflicts $R_k$ |
| *Delete $R_i$* | Delete $R_k$ & Delete relation | Delete $R_k$ & Delete relation | Delete property of $R_k$ | Delete relation or (Delete $R_k$ & Delete relation) | Delete relation |
| *Add property to $R_i$* | No impact | Add property to $R_k$ or Delete relation | Delete relation | No impact | No impact |
| *Delete property of $R_i$* | No impact or Delete relation or (Delete $R_k$ & Delete relation) or Delete property of $R_k$ | Delete property of $R_k$ or (Delete property of $R_k$ & Delete relation) | Delete property of $R_k$ | No impact or Delete relation or (Delete $R_k$ & Delete relation) | No impact or Delete relation |
| *Add constraint to property of $R_i$* | No impact or Add constraint to property of $R_k$ or Delete relation | No impact | No impact | No impact | No impact |
| *Delete constraint of property of $R_i$* | No impact or Delete constraint of property of $R_k$ | No impact or Delete relation or Delete constraint of property of $R_k$ or (Delete constraint of property of $R_k$ & Delete relation) | No impact or Delete relation or Delete constraint of property of $R_k$ or (Delete constraint of property of $R_k$ & Delete relation) | No impact or Delete relation or (Delete $R_k$ & Delete relation) | No impact or Delete relation |

Consider the following requirements for the RPM system:

**R1** *The system shall measure temperature from a patient.*

**R4** *The system shall store patient temperature measured by the sensor in the central storage.*

**R6** *The system shall store data measured by sensors in the central storage.*

**R7** *The system shall warn the doctor when the temperature threshold is violated.*

**R8** *The system shall generate an alarm if the temperature threshold is violated.*

**R9** *The system shall show the doctor the temperature alarm at the doctors' computers.*

**R14** *The system shall store patient temperature measured by the sensor in the central storage and it shall warn the doctor when the temperature threshold is violated.*

The stakeholder poses a change for the RPM system: The system shall warn the doctor with all information about the patient's condition (blood pressure, temperature, etc.) when the temperature threshold is violated. One of the properties in R14 is "warning the doctor when the temperature threshold is violated". We propose the following change.

**Proposed Change:** Add Constraint to Property of R14

**Description of the Proposed Change:** The warning to the doctor should also contain all information about the patient's condition.

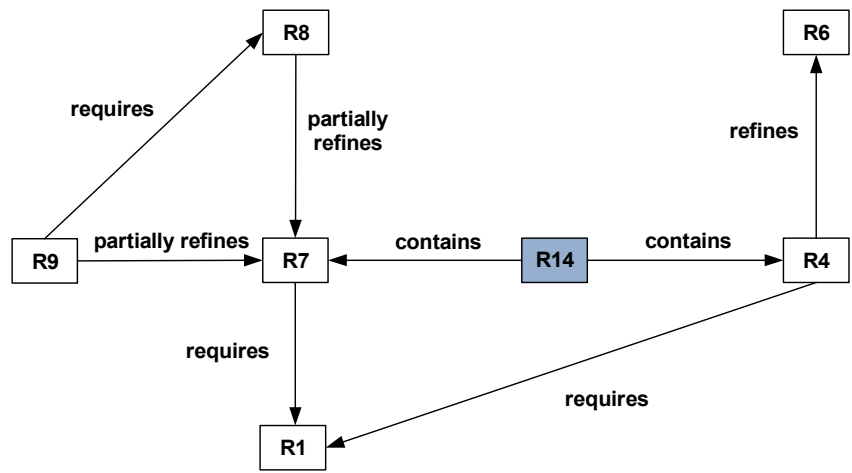

Figure 7 Requirements Related to R14 with Distance of 2

The proposed change is propagated to the requirements related to R14. Figure 7 gives the requirements related to R14 with distance of 2 (inferred relations are not shown for simplicity). The distance is the number of relations between two requirements.

TRIC automatically lists the impact alternatives for the requirements R7 and R4 which are directly related to R14. R14 contains the properties (*storing patient temperature* and *warning the doctor*) given in R4 and R7 via the *contains* relation. There are two change alternatives to propagate the change from R14 to R7 and to R4 via the *contains* relation: 'Add Constraint to Property of Requirement' or 'No Impact' (see Table 3). The change type 'Add Constraint to Property of Requirement' is chosen for R7 since the property in R7 is the impacted property in R14. 'No Impact' is chosen for R4 since it gives the property "Storing patient temperature" which does not have anything related to the added constraint.

**Proposed Change for R7:** Add Constraint to Property of R7

**Description of the Proposed Change:** The warning to the doctor should also contain all information about the patient's condition.

The next step is to propagate the change to the indirectly related requirements. Since 'No Impact' is chosen for R4, we do not have to check R6 indirectly related to R14 via R4 (see Figure 7). Therefore, the next propagation is from R7 to its related requirements. We do not illustrate the next step for brevity but the details of the approach and a complete example can be found in our previous work [20].

The requirements engineer selects among the provided alternatives to propagate the proposed change from one requirement to another step-by-step. As an output of the propagation process we have a set of proposed changes for the impacted requirements with a propagation path in the requirements model.

## 7 Identifying Impacted Architectural Elements

The approach in Section 6 enables the requirements engineer to propose a change for a requirement and to propagate the proposed change to related requirements. The output is a set of proposed/propagated requirements changes in a propagation path in the requirements model. In this paper our technique focuses on determining architectural elements that implement system properties given in requirements to which changes are proposed/propagated. We are concerned with *domain changes* for requirements (see Section 5.2). By using the formal semantics of requirements relations, changes and traces between R&A, we identify which parts of software architecture are impacted by a proposed requirements change. The impact is calculated by a change impact function. The function takes a change type, a requirement to which the change is introduced, a set of requirements relations for the requirement and a set of all traces between R&A, as input. The output is a set of architectural elements which are candidate to be impacted by the requirements change. The following is the signature of the change impact function.

*impact* : *SCT* × *SR* × *SSRR* × *SST* → *SSAE*

where *SCT* is the set of change types, *SR* is the set of requirements, *SSRR* is the set of sets of requirements relations, *SST* is the set of sets of traces and *SSAE* is the set of sets of architectural elements which are candidate in the impact for the requirements change.

Traces in *SST* can be either generated *Satisfies* or assigned *AllocatedTo* traces. Given the domain changes that can be made to the requirements model, we describe change impact rules to determine the impact of each requirements change type in the software architecture. All change impact rules are derived from the semantics of requirements, requirements relations, changes and traces between R&A. According to the type of the change and relations in the propagation path, the function may traverse the propagation path in the requirements model. It identifies which impacted requirement(s) in the propagation path should be traced to determine candidate architectural elements. Then, candidate architectural elements for the impact are identified by tracing from requirements to architecture.

- **Candidate Architectural Elements for 'Add a New Requirements Relation', 'Delete Requirements Relation', and 'Update Requirements Relation':** There is no impact on the architecture. These changes do not modify any system property in the requirements model (see *refactoring*).

- **Candidate Architectural Elements for 'Add Property to Requirement':** Our approach returns no suggestion for candidate architectural elements. The software architect needs to manually analyze the change and impacted requirement(s) to identify candidate architectural elements. If the added property is a new system property (see *domain changes*), architectural elements that satisfy the existing properties related to the added property are candidate for the impact. In the requirement itself, there is no explicit dependency between the existing properties and the added property. Therefore, it is not possible to automatically identify architectural elements that satisfy the existing properties related to the added property. The added property may be an existing property added to the requirement to improve the structure of the model without modifying overall system properties (see *refactoring*). In this case there is no impact on software architecture. We cannot identify automatically if the added property is a new system property.

- **Candidate Architectural Elements for 'Add a New Requirement':** Either architectural elements traced from directly related requirements are candidate for the impact or there is no impact (see Section 7.1).

- **Candidate Architectural Elements for 'Delete Requirement' and 'Update Requirement':** The propagation path of the change is traversed to identify candidate architectural elements for the impact (see Section 7.2).

Some candidate architectural elements may not be actually impacted because of design decisions taken by the software architect. New architectural elements might be introduced instead of changing the existing ones. Candidate architectural elements are input to take design decisions in the implementation of the change.

## 7.1 Candidate Architectural Elements for 'Add a New Requirement'

An added requirement may introduce a new system property (*domain changes*). In this case, architectural elements that satisfy existing requirements directly related to the added requirement are candidate for the impact. If there is no new system property (*refactoring*), there is no impact on architecture. When there is no dependency between the existing requirements and the added requirement, it is not possible to automatically identify any candidate architectural element. Table 4 gives the impact rules for 'Add a New Requirement'.

**Table 4** Change Impact Rules for the Change Type 'Add a New Requirement'

| Change | Requirements Relation Types | | | | | | | |
|---|---|---|---|---|---|---|---|---|
| | $R_i$ contains $R_x$ | $R_i$ refines $R_x$ | $R_i$ partially refines $R_x$ | $R_i$ requires $R_x$ | $R_x$ contains $R_i$ | $R_x$ refines $R_i$ | $R_x$ partially refines $R_i$ | $R_x$ requires $R_i$ |
| Add $R_x$ | No impacted AE[3] | No impacted AE | No impacted AE | AEs traced from $R_i$ are candidate | No impacted AE | AEs traced from $R_i$ are candidate | AEs traced from $R_i$ are candidate | AEs traced from $R_i$ are candidate |

Each cell gives candidate architectural elements for the change type in the row and the relations in the columns. $R_i$ and $R_x$ denote existing and added requirements respectively. 'Add a New Requirement' is not a domain change if ($R_i$ *contains* $R_x$), ($R_i$ *refines* $R_x$), ($R_i$ *partially refines* $R_x$) or ($R_x$ *contains* $R_i$). Therefore, there is no impact on architecture.

Please note that all impact rules are derived from the semantics of requirements, requirements relations, changes and traces between R&A. The following is the justification of the change impact rule in Table 4 for the change 'Add a New Requirement' (Add $R_x$) where ($R_x$ refines $R_i$).

**Change Impact Rule for 'Add a New Requirement' (Add $R_x$) where ($R_x$ refines $R_i$)**

Candidate architectural elements for the change type 'Add a New Requirement' (Add $R_x$) where ($R_x$ refines $R_i$)

= Architectural elements traced from $R_i$ are candidate

---

[3] 'AE' stands for 'Architectural Element'

**Justification:**

Let $R_i$, $R_x$ be requirements and $E_A$ be the set of architectural elements that *satisfies* $R_i$ where $P_i$ and $P_x$ are formulas for $R_i$ and $R_x$, and $P_A$ is the formula for the system property $E_A$ is needed to implement.

= {*By using formalization of the refines relation*}

$P_x \rightarrow P_i$

= {*By using formalization of the satisfies trace*}

The fulfillment of $P_A$ implies the fulfillment of $P_i$

$P_i$ also holds for the set of architectural elements $E_A$. The new requirement $R_x$ is a refinement of $R_i$. Usually, the architectural elements in $E_A$ provide part of the functionality that satisfies $P_x$. The elements in $E_A$ can be reused or adapted in order to implement the new requirement $R_x$. Therefore, they are candidate architectural elements for the impact.

The following is the justification of the rule in Table 4 where the change 'Add a New Requirement' is not a domain change (Add $R_x$ where $R_i$ *contains* $R_x$).

**Change Impact Rule for 'Add a New Requirement' (Add $R_x$) where ($R_i$ contains $R_x$)**

Candidate architectural elements for the change type 'Add a New Requirement' (Add $R_x$) where ($R_i$ contains $R_x$)

= No candidate architectural element

**Justification:**

Let $RM$ be a requirements model where $P_{RM}$ is the formula for $RM$.

The requirements model $RM$ is the set of requirements $R_1, R_2, \ldots, R_k$ where $P_1, P_2, \ldots, P_k$ are formulas for $R_1, R_2, \ldots, R_k$, and $k \geq 1$. $P_{RM}$ can also be represented in the following way:

$P_{RM} = P_1 \wedge P_2 \wedge \ldots \wedge P_k$

Please note that if the requirements $R_1, R_2, \ldots, R_k$ are written as formulas $\forall x \varphi_1, \forall x \varphi_2, \ldots, \forall x \varphi_k$ with $\varphi_1, \varphi_2, \ldots, \varphi_k$ in CNF, we have the following: ($P_{RM} = \forall x (\varphi_1 \wedge \varphi_2 \wedge \varphi_3 \wedge \ldots \wedge \varphi_k)$).

Let $R_i$ and $R_x$ be requirements where $P_i$ and $P_x$ are formulas for $R_i$ and $R_x$, and (i $\leq$ k)

Let $RM^l$ be the requirements model after the change 'Add $R_x$' where $P_{RM}{}^l$ is the formula for $RM^l$.

= {*By using formalization of the change type 'Add Requirement'*}

$P_{RM}{}^l = P_{RM} \land P_x$

If $P_{RM}$ and $P_x$ are written as formulas $\forall x(\varphi_1 \land \varphi_2 \land \varphi_3 \land \ldots \land \varphi_k)$ and $\forall x \varphi_x$ with $\varphi_1, \varphi_2, \ldots, \varphi_k, \varphi_x$ in CNF, we have the following: ($P_{RM}{}^l = \forall x(\varphi_1 \land \varphi_2 \land \varphi_3 \land \ldots \land \varphi_k \land \varphi_x)$).

= {*By using formalization of the contains relation*}

We have the following: ($P_i = P_x \land P^l$) where $P^l$ denotes properties that are not captured in $P_x$. Please note that if the requirements $R_i$ and $R_x$ are written as formulas $\forall x \varphi_i$ and $\forall x \varphi_x$ with $\varphi_i$ and $\varphi_x$ in CNF and $P^l$ is expressed as $\forall x \psi$ with $\psi$ in CNF, we understand the following: $R_i$ *contains* $R_x$ iff ($P_i = \forall x(\varphi_x \land \psi)$), and ($\neg(\forall x(\varphi_x \to \varphi_i))$) and ($\neg(\forall x(\psi \to \varphi_i))$) are satisfiable.

$P_{RM}{}^l = P_{RM} \land P_x$

$P_{RM}{}^l = \forall x(\varphi_1 \land \varphi_2 \land \varphi_3 \land \ldots \land \varphi_k \land \varphi_x)$

$P_{RM}{}^l = \forall x(\varphi_1 \land \varphi_2 \land \ldots \land \varphi_i \land \ldots \land \varphi_k \land \varphi_x)$

$P_{RM}{}^l = \forall x(\varphi_1 \land \varphi_2 \land \ldots \land \varphi_x \land \psi \land \ldots \land \varphi_k \land \varphi_x)$

$P_{RM}{}^l = \forall x(\varphi_1 \land \varphi_2 \land \ldots \land \varphi_x \land \psi \land \ldots \land \varphi_k)$

$P_{RM}{}^l = \forall x(\varphi_1 \land \varphi_2 \land \ldots \land \varphi_i \land \ldots \land \varphi_k)$

$P_{RM}{}^l = \forall x(\varphi_1 \land \varphi_2 \land \varphi_3 \land \ldots \land \varphi_k)$

Then we get $P_{RM}{}^l = P_{RM} = \forall x(\varphi_1 \land \varphi_2 \land \varphi_3 \land \ldots \land \varphi_k)$

= {*By using the formalization of domain changes and refactoring*}

Properties that are captured in the requirements model $RM$ are preserved in the new requirements model $RM^l$ and there is no new property in the new requirements model $RM^l$. Therefore, we can conclude that the architecture, that satisfies requirements in the requirements model $RM$, satisfies requirements in the requirements model $RM^l$ after the change 'Add $R_x$'. There is no need to change the architecture and therefore, there is no candidate architectural element.

The following is a change impact example with 'Add a New Requirement' in the RPM system.

**Change Impact Example with 'Add a New Requirement' (Add $R_x$)**

We explain one of the change impact rules in Table 4 for the change type 'Add a New Requirement' with the following requirements of the RPM system.

**R5** *The system shall store patient blood pressure measured by the sensor in the central storage.*

**R15** *The system shall store patient Central Venous Pressure (CV Pressure) measured by the sensor in the central storage.*

where (R15 *refines* R5)

The stakeholders need the following change: Measuring and storing blood pressure is refined further for Pulmonary Artery Pressure (PA pressure). Therefore, we propose the change 'Add a New Requirement' and the new requirement refines R5.

**RX** *The system shall store patient Pulmonary Artery Pressure (PA pressure) measured by the sensor in the central storage.*

where (RX *refines* R5)

According to Table 4, architectural elements traced from R5 are candidate for the impact of adding RX where RX refines R5. Figure 8 shows the *Satisfies* traces for R5 and R15 with the candidate architectural elements.

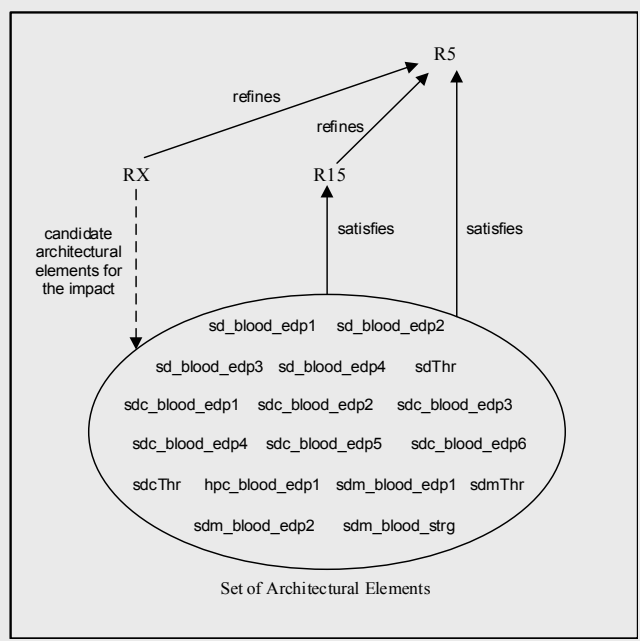

**Figure 8** Candidate Architectural Elements for the Added Requirement

Since the property in RX is the refinement of the system property given in R5, the architectural elements that implement storing patient blood pressure are candidate to be impacted for storing patient PA pressure. Figure 9 shows the part of the RPM architecture in AADL visual notation, that satisfies R5.

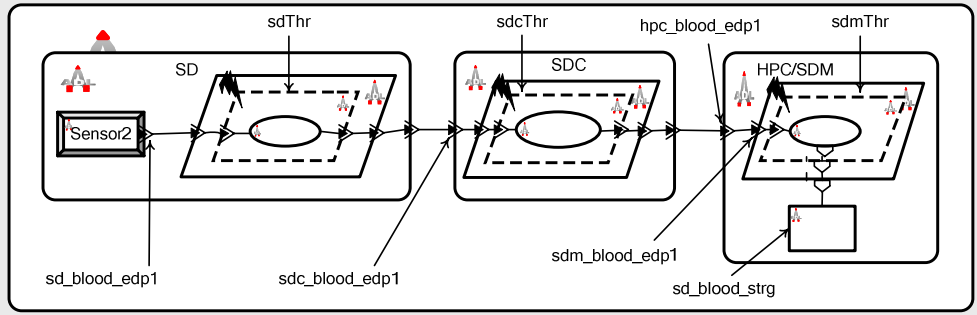

**Figure 9** Part of the RPM Architecture for Storing Blood Pressure

Before adding RX, R15 is the only requirement that refines R5. Therefore, the part of the RPM architecture in Figure 9 satisfies the refined property in R15 (*Storing patient CV pressure measured by the sensor*). We inspected the architecture based on the new requirement and the candidate architectural elements. We changed the architecture to get the new requirement satisfied by the architecture. Figure 10 gives the changed part of the RPM architecture.

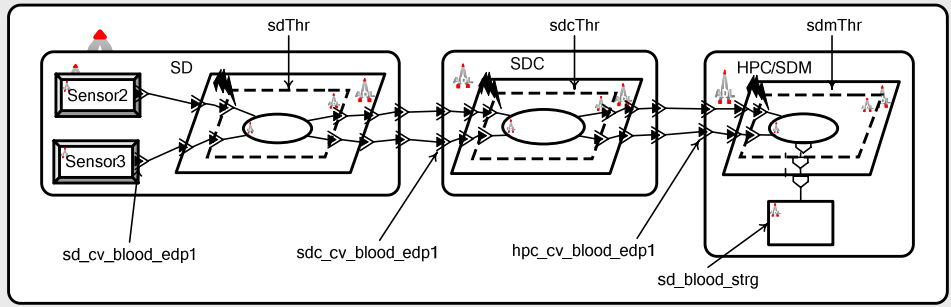

**Figure 10** Changed Part of the RPM Architecture for Pulmonary Artery Pressure

We added a new sensor (*Sensor 3*) and new event data ports (*sd_pa_blood_edp1*, *sdc_pa_blood_edp1*, etc.) to store the patient PA pressure. The threads *sdThr*, *sdcThr* and *sdmThr* have some new event data ports. The measured PA pressure is stored in the existing data store (*sd_blood_strg*). According to our changes, the actual impacted architectural elements are the threads *sdThr*, *sdcThr*, *sdmThr* and the data store *sd_blood_strg*.

The candidate architectural elements might not have been actually impacted at all. For instance, we could have proposed new event data ports, threads, sensors, and data storages. None of the candidate architectural elements would have been affected. With candidate elements, we aim at identifying architectural elements that satisfy existing properties related to the new property. To implement the new property in the added requirement, the software architect is guided with the architectural elements that are potentially related to the new property.

## 7.2 Candidate Architectural Elements for 'Delete Requirement' and 'Update Requirement'

For 'Delete Requirement' and 'Update Requirement' (except 'Add Property to Requirement') as *domain* changes, architectural elements satisfying changed properties are candidate for the impact. Changed requirements may also have unchanged properties. The propagation path needs to be traversed to identify impacted requirement(s) which has the least number of unimpacted properties.

We define a recursive function traversing the propagation path. The function takes a change, a requirement to which the change is introduced, a set of relations of the requirement in the propagation path, and a set of traces between R&A as input. The output is a set of candidate architectural elements for the impact.

**Table 5** Traversal Rules for the Change Types "Delete Requirement" and "Update Requirement"

| Changes | Requirements Relation Types | | | | | |
|---|---|---|---|---|---|---|
| | $R_i$ contains $R_k$ | $R_i$ refines $R_k$ | $R_i$ partially refines $R_k$ | $R_k$ contains $R_i$ | $R_k$ refines $R_i$ | $R_k$ partially refines $R_i$ |
| Delete $R_i$ | Do not take $R_k$ in the traversal | Do not take $R_k$ in the traversal | Do not take $R_k$ in the traversal | Do not take $R_k$ in the traversal | Take $R_k$ in the traversal | Do not take $R_k$ in the traversal |
| Delete Property of $R_i$ | Take $R_k$ in the traversal | Do not take $R_k$ in the traversal | Do not take $R_k$ in the traversal | Do not take $R_k$ in the traversal | Take $R_k$ in the traversal | Take $R_k$ in the traversal |
| Change Property of $R_i$ | Take $R_k$ in the traversal | Do not take $R_k$ in the traversal | Do not take $R_k$ in the traversal | Do not take $R_k$ in the traversal | Take $R_k$ in the traversal | Take $R_k$ in the traversal |
| Add Constraint to Property of $R_i$ | Take $R_k$ in the traversal | Do not take $R_k$ in the traversal | Do not take $R_k$ in the traversal | Do not take $R_k$ in the traversal | Take $R_k$ in the traversal | Take $R_k$ in the traversal |
| Delete Constraint of Property of $R_i$ | Take $R_k$ in the traversal | Do not take $R_k$ in the traversal | Do not take $R_k$ in the traversal | Do not take $R_k$ in the traversal | Take $R_k$ in the traversal | Take $R_k$ in the traversal |
| Change Constraint of Property of $R_i$ | Take $R_k$ in the traversal | Do not take $R_k$ in the traversal | Do not take $R_k$ in the traversal | Do not take $R_k$ in the traversal | Take $R_k$ in the traversal | Take $R_k$ in the traversal |

The function traverses the propagation path based on the traversal rules in Table 5 until there is no more relation to be taken. The reader is referred to Supplementary Material C[4] for the algorithm of the function.

---

[4] http://people.svv.lu/goknil/supplementary/SupplementaryMaterialC.pdf

Table 5 has change types in the rows and requirements relation types in the columns. The traversal rules are derived from the semantics of the change and relation types. A relation in the column is considered for a change in the corresponding row only if the relation is in the propagation path. The changes in the rows of Table 5 represent the changes selected by the requirements engineer among the alternatives in the change impact alternative table (see Section 6).

In the following we explain the traversal function with a simple model. Figure 11 gives an example requirements model where 'Change Constraint of Property of Requirement' is proposed to R2 and propagated to R1 and R5.

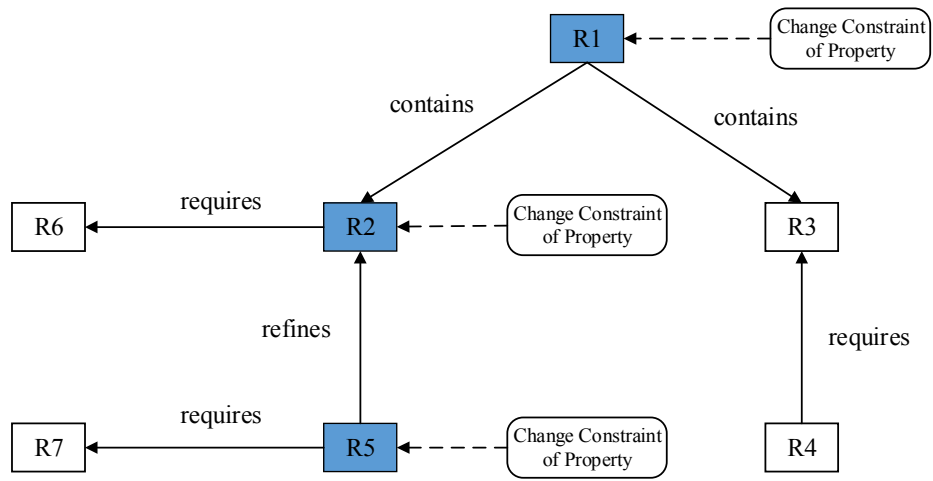

**Figure 11** Example Requirements Model with Impacted Requirements

The requirements engineer can select any impacted requirement in the propagation path to identify candidate architectural elements:

- R5 is selected. In the propagation path, R5 refines R2. The function does not take R2 in the traversal of the path (see the cell for '*Change Constraint of Property of $R_i$*' and '*$R_i$ refines $R_k$*' in Table 5). There is no more relation of R5 in the propagation path. Therefore, architectural elements satisfying R5 are candidate for the impact in the architecture.

- R2 is selected. In the propagation path, R1 contains R2 and R5 refines R2. The function does not take R1 in the traversal (see the cell for '*Change Constraint of Property of $R_i$*' and '*$R_k$ contains $R_i$*'). It takes R5 to traverse the path (see the cell for '*Change Constraint of Property* of $R_i$' and '*$R_k$ refines $R_i$*'). There is no more relation of R2 and R5 in the propagation path. Therefore, architectural elements satisfying R5 are candidate.

- R1 is selected. In the propagation path, R1 contains R2. The function takes R2 to traverse the path (see the cell for 'Change Constraint of Property of $R_i$' and '$R_i$ contains $R_k$'). In the path, R5 refines R2. As a recursive call, the function takes R5 to traverse the path (see the cell for 'Change Constraint of Property of $R_i$' and '$R_k$ refines $R_i$'). There is no relation of R5 in the path. Architectural elements satisfying R5 are candidate.

The following is the justification of the traversal rule in Table 5 for the change 'Add Constraint to Property of Requirement' for $R_i$ where ($R_k$ partially refines $R_i$).

**Traversal Rule for the Change 'Add Constraint to Property of Requirement'**

Candidate architectural elements for the change type 'Add Constraint to Property of Requirement' for $R_i$ where ($R_k$ partially refines $R_i$) and the change is propagated to $R_k$ (Add Constraint to Property of Requirement $R_k$)

$\quad\quad\quad$ = Take $R_k$ to traverse the propagation path

**Justification:**

Let $R_i$ be a requirement where $P_i$ is the formula for $R_i$. $P_i$ is represented in a conjunctive normal form (CNF) in the following way:

$\quad P_i = \forall x\,(p_1 \ldots p_n);\ n \geq 1$ and $p_i$ is disjunction of literals

Let $R_k$ be a requirement where $P_k$ is the formula for $R_k$.

Let $R_i^1$ and $R_k^1$ be the requirements after the changes (Add Constraint to Property of Requirement $R_i$) and (Add Constraint to Property of Requirement $R_k$) where $P_i^1$ and $P_k^1$ are the formulas for $R_i^1$ and $R_k^1$.

Let $E_{Ai}$ be the set of architectural elements that *satisfies* $R_i$ and $E_{Ak}$ be the set of architectural elements that *satisfies* $R_k$ where $P_{Ai}$ is the formula for the system property $E_{Ai}$ is needed to implement and $P_{Ak}$ is the formula for the system property $E_{Ak}$ is needed to implement.

= {*By using formalization of the satisfies trace*}

$\quad$ The fulfillment of $P_{Ai}$ implies the fulfillment of $P_i$

$\quad$ The fulfillment of $P_{Ak}$ implies the fulfillment of $P_k$

= {*By using formalization of the partially refines relation*}

$\quad P_k = \forall x\,(p_1^1 \ldots p_z^1);\ z < n$ and $\forall x\,(p_j^1 \to p_j)$ for for all $j \in 1..z$

= {*By using formalization of the change type 'Add Constraint to Property of Requirement' for $R_i$*}

$$P_i^1 = \forall x \, ((p_1^{ll} \ldots p_t^{ll}) \wedge (p_{t+1} \ldots p_n)); \, t \leq z \text{ and } \forall x \, (p_j^{ll} \rightarrow p_j) \text{ for all } j \in 1..t$$

The properties captured in $\forall x \, (p_{z+1} \ldots p_n)$ in $R_i$ are not affected by the change. These properties are not captured by $R_k$. Therefore, the propagation path is traversed for $R_k$.

The following is a change impact analysis example with the change 'Add Constraint to Property of Requirement' in the RPM system.

**Change Impact Example with 'Add Constraint to Property of Requirement'**

We explain the impact analysis for the change type 'Add Constraint to Property of Requirement' with the example change propagation given in Section 6.

The stakeholders require a change for the RPM system: The system shall warn the doctor with the patient's condition information when the temperature threshold is violated. Initially, the change 'Add Constraint to Property of Requirement' is proposed for R14.

**R14** *The system shall store patient temperature measured by the sensor in the central storage and it shall warn the doctor when the temperature threshold is violated.*

**Proposed Change:** Add Constraint to Property of R14

**Description of the Change:** If the temperature threshold is violated, the system shall warn the doctor with the patient's condition information.

The proposed change is propagated in the requirements model (see Figure 12).

**R4** *The system shall store patient temperature measured by the sensor in the central storage.*

**R7** *The system shall warn the doctor when the temperature threshold is violated.*

**R8** *The system shall generate an alarm if the temperature threshold is violated.*

**R9** *The system shall show the doctor the temperature alarm at the doctors' computers.*

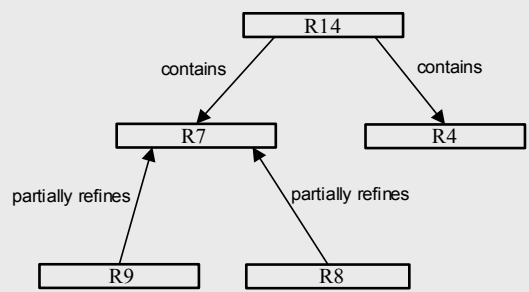

**Figure 12** Part of the RPM Requirements Model

The property of R14 is 'warning the doctor about the temperature threshold violation'. The constraint added to the property of R14 is 'warning the doctor with the patient's condition information'. The proposed change is propagated to the requirements which contain or refine the property 'warning the doctor about the temperature threshold violation' (see Figure 13 for the propagation path).

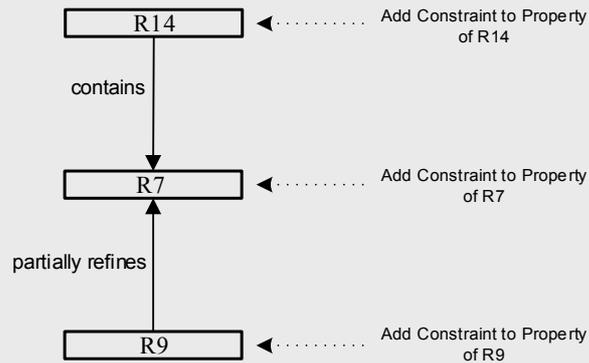

**Figure 13** Propagation Path of the Proposed Change in R14

The changes for R7 and R9 in the propagation path are the following:

**Proposed Change for R7:** Add Constraint to Property of R7

**Description of the Proposed Change:** If the temperature threshold is violated, the system shall warn the doctor with the patient's condition information.

**Proposed Change for R9:** Add Constraint to Property of R9

**Description of the Proposed Change:** The system shall show the doctor the temperature alarm with the patient's condition information at the doctor's computer.

The requirements engineer can select any impacted requirement in the propagation path to identify the candidate architectural elements. We assume that R14 is selected by the software architect. To identify the candidate architectural elements, the change impact function traverses the propagation path in Figure 13 based on the traversal rules in Table 5.

According to Table 5, $R_k$ is taken to traverse the propagation path when the change is 'Add constraint to property of $R_i$' and ($R_i$ *contains* $R_k$). Since R14 has the change 'Add Constraint to Property of Requirement' and R14 *contains* R7, R7 is taken to traverse the propagation path.

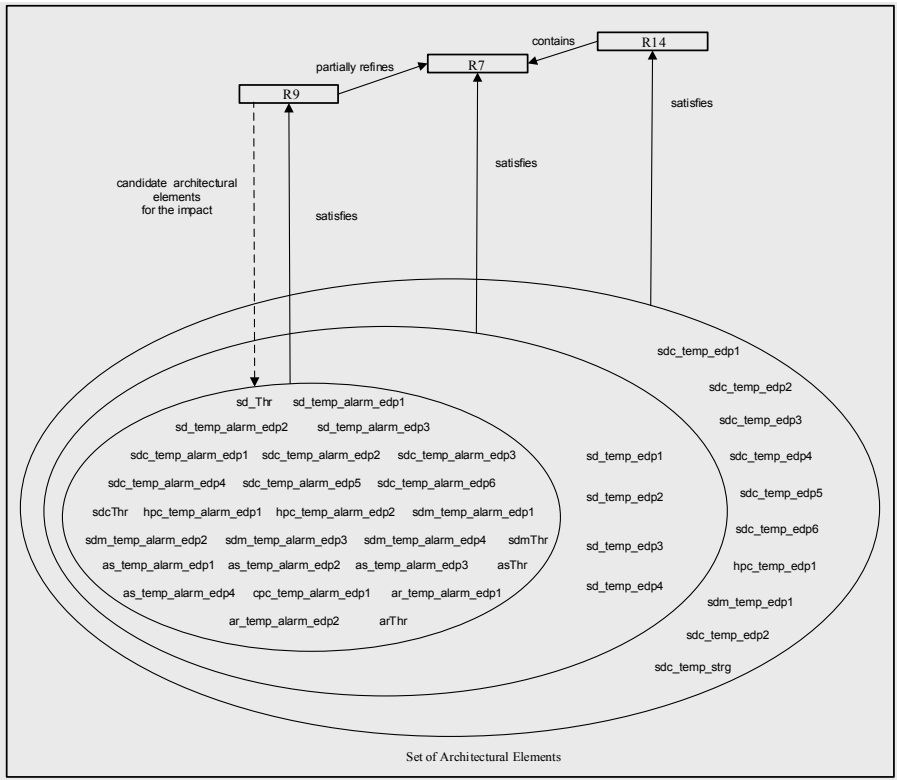

**Figure 14** Candidate Architectural Elements for the Constraint Added to R14

R9 is taken to traverse the propagation path as a next step since R9 has the change 'Add Constraint to Property of Requirement' and R9 *partially refines* R7 (see the cell for 'Add constraint to property of $R_i$' and '$R_k$ *partially refines* $R_i$' in Table 5). There is no other requirement in the propagation path. Therefore, the architectural elements traced from R9 are candidate for the impact (see Figure 14).

R9 has the most refined property impacted by the proposed change. Therefore, the architectural elements satisfying R9 are identified as candidate to implement the changes proposed for R7, R9, and R14. Figure 15 shows the part of the RPM architecture that satisfies R9.

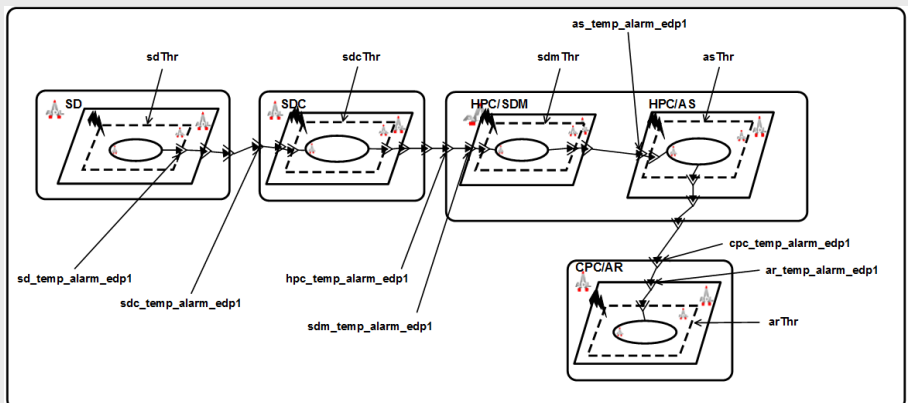

**Figure 15** Part of the RPM Architecture for Showing the Temperature Alarm

When the temperature threshold is violated, the *SD* component generates an alarm and sends it through the *SDC* component and the *SDM & AS* subcomponents in the *HPC* component to the *AR* subcomponent in the *CPC* component. The *AR* notifies the doctor of the alarm. Based on the candidate elements, we changed the architecture to get the new constraint satisfied by the architecture (see Figure 16).

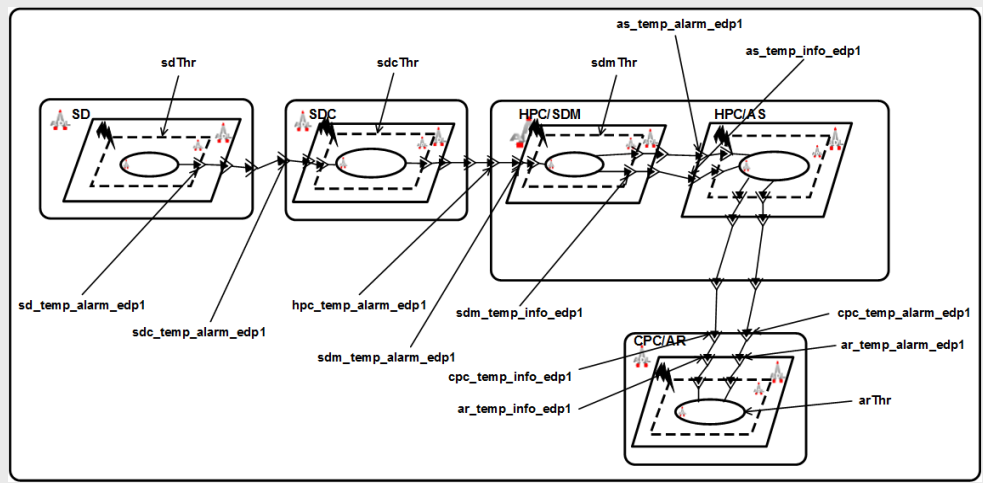

**Figure 16** Changed Part of the RPM Architecture for the Changed Requirements

The *SDM* has the access to the information about the patient's condition such as temperature information. For the temperature information we added new event data ports (*sdm_temp_info_edp1*, *as_temp_info_edp1*, *cpc_temp_info_edp1*, etc.) starting from the *SDM* to the *AR* through the *AS*. Please note that more data ports should be added to the architecture for other information such as blood pressure. For brevity we only included the event data ports for the temperature information. When the generated alarm is received by the *SDM*, the *sdmThr* thread transmits the patient's temperature information to the *AR* via the newly added event data ports. The *AR* shows the generated alarm with the temperature information to the doctor. To enable the transmission, we updated the dynamic behaviour of the *sdmThr*, *asThr* and *arThr* threads in the behavioral annex of the RPM architecture.

## 8  Tool Support

The tool support is the combination of TRIC and Eclipse Model Editor. Figure 17 gives the GUI to select the proposed requirements change in TRIC.

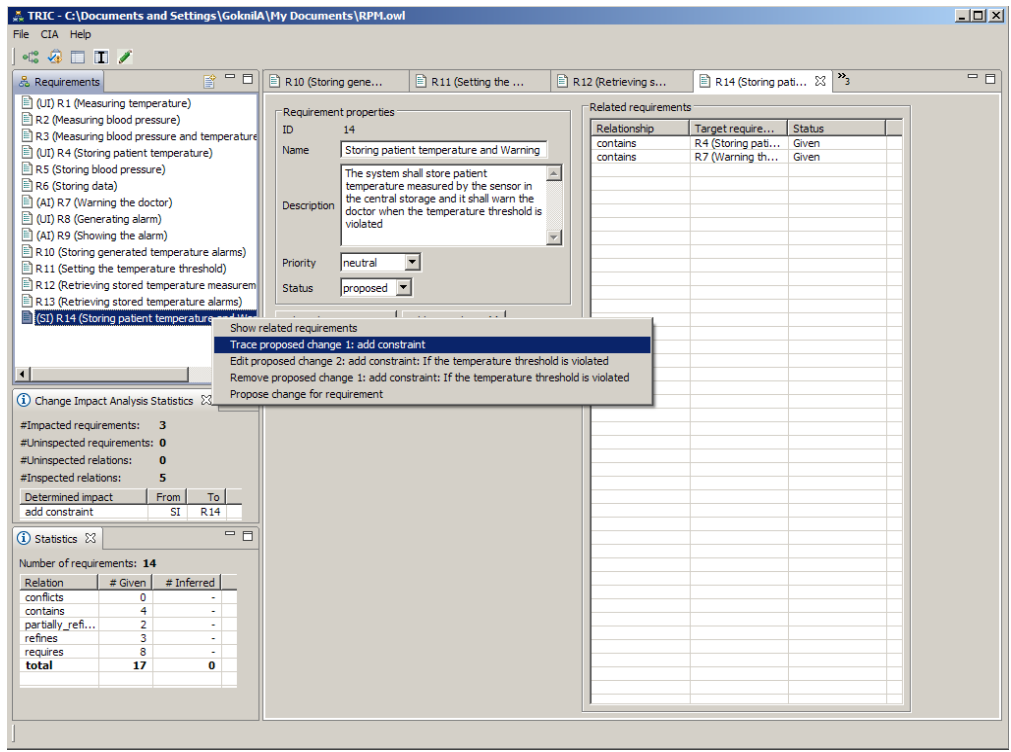

**Figure 17** GUI for Selecting the Proposed Requirements Change in TRIC

The left-hand side of the window lists the requirements in the requirements model. The requirements are tagged as *SI (Starting Impacted)* and *UI (Unimpacted)*. The right-hand side shows the details of the selected requirement (R14). The pop-up menu is used to select a proposed change in the selected requirement. The propagation path is traversed (if needed) starting from the selected requirement and change. Figure 18 gives the output of the traversal for the change 'Add Constraint to Property of Requirement' in R14 (see Section 7.2).

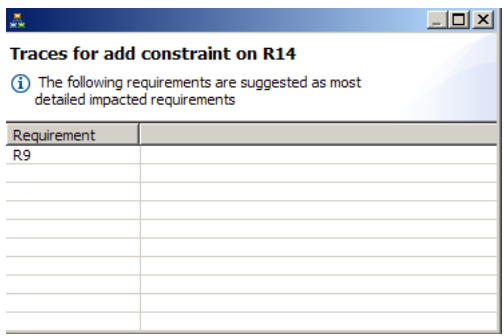

**Figure 18** Output of Traversing the Propagation Path of the Proposed Change in R14

Architectural elements traced from the impacted requirement(s) in Figure 18 are candidate for the impact of the selected change in Figure 17. The Eclipse Model Editor is used to trace from the impacted requirement(s) to candidate architectural elements in the trace model.

## 9 Discussion of the Approach

**Chosen Formalization.** In our approach, we use a formalization of requirements, their relations and change types in First-Order Logic (FOL). There are other formalizations of requirements, for example, in modal logic and deontic logic [37]. With the formalization in FOL, we can express commonly occurring requirement descriptions such as real-time and performance requirements. However, there are limitations of the expressivity of FOL. For instance, we do not cover imperfect requirements which can be modeled by fuzzy sets [41]. Dealing with modalities in requirements such as possibility, probability, and necessity is out of scope of our formalization. Our formalization needs to be extended with temporal logic, modal logic or fuzzy sets in order to cover these types of requirements and their changes. Under these limitations, the expressiveness of FOL is sufficient for our change impact analysis approach since our focus is on the commonly occurring requirements.

**Chosen Metamodels.** Our requirements metamodel contains common entities identified in the literature for requirements models. In the literature there are multiple requirements engineering practices and methods using different structuring and definition of requirements. For instance, Goal-oriented requirements engineering [54] [40] provides a model which allows for decomposing a system goal into requirements with goal trees. The variability management approach [38] deals with producing requirements that can be considered as a core asset in a product line. Since the focus of our approach is on the commonly occurring requirements relations, we investigated and benefited from all these requirements engineering methods which are commonly used to define and represent requirements. The main criterion for constructing our metamodel is the most commonly occurring requirements relations. In the metamodel we left out other entities such as goals, stakeholders, and test cases. The selected relation types in our requirements metamodel are compatible with the results of an industrial case study [58] which evaluates the applicability of existing dependency types in the literature.

In our previous work [19], we also show how our requirements metamodel can be customized for different requirements modeling approaches and notations such as Product-line and SysML. Mainly, the requirements relations in our metamodel are customized to support relations in different forms of requirements.

The customization allows using the same semantics and reasoning mechanism of our requirements metamodel for multiple forms of requirements.

We do not have an explicit separation of functional and non-functional requirements in the requirements metamodel. However, the *Requirements* entity and the requirements relations in the metamodel are sufficient to model both types of requirements. For instance, for a performance requirement *PR* of a system functionality given in a functional requirement *FR*, it might be the case that *PR* refines *FR*. In our previous work [24] [22] [20] we show how we can model both functional and non-functional requirements with their dependencies based on our requirements metamodel.

The literature also proposes several types of traces, which are similar to the trace types in our trace metamodel but named differently. Paige et al. [45] classifies trace types between any design models and informal requirements as *consistent-with*, *dependency*, *satisfies*, *allocated-to*, and *refines*. Khan et al. [32] propose six types of traces which differ only in the type of the source requirement. Our trace metamodel is abstracted to keep only the very generic types *Satisfies* and *AllocatedTo*.

**Generalization of Our Approach.** In this paper we use AADL to model the RPM architecture. On the other hand, software architecture can be expressed with any other ADL or more generic notations like UML. We only require traces between requirements and software architecture. Our approach identifies which impacted requirement(s) should be traced to software architecture to determine candidate architectural elements. In this respect, our approach is independent of any notation and tool used for modeling software architecture.

**Maintaining Traces.** The need of trace maintenance depends on the rationale of requirements changes. *Domain changes* modify the overall system properties and require changes in the architecture. With domain changes, traces between changed requirements and (un-)changed architectural elements might be invalid as well as traces between unchanged requirements and changed architectural elements. *Refactoring changes* does not modify the overall system properties. Therefore, only traces from changed requirements might be invalid in refactoring changes.

**Tracing Requirements and Architecture to Source Code.** Our approach can be combined with code-based change impact analysis techniques [35] using traces between architecture and source code. In addition, some trace generation methods and tools [25] [28] [39] can be utilized to establish traces between requirements/architecture and source code. If source code is automatically generated from the architecture, code changes can be traced and implemented automatically. In some cases, requirements changes are traced directly to source code. It is still beneficial to keep the architecture synchronized with the code and to trace the changes from requirements to architecture because architecture is an aid in understanding and communicating the design of complex systems.

## 10 Related Work

A number of approaches in the literature address change impact analysis in software architecture. Jonsson and Lindvall [31] present common strategies for change impact analysis in two categories: *automatable* (traceability/dependency analysis and slicing techniques) and *manual* (design documentation and interviews). Automatable analysis strategies often employ algorithmic methods for change propagation [31]. Traceability analysis is the analysis of any relation among artifacts. Therefore, our approach can be considered traceability analysis.

Algorithmic methods are employed by Lee et al. [34] to compute the impact of changes on object-oriented software. Lee et al. uses data dependency graphs with a classification of changes for object-oriented software to determine the impacted elements in object-oriented source code. The approach addresses the impact analysis in source code, not in high-level design. Tang et al. [50] introduce Architecture Rationale and Element Linkage (AREL) model represented as a Bayesian Belief Network (BBN). AREL captures the casual relationship between architectural elements and decisions using propabilities. This allows architects to perform change impact in software architecture based on probability theory. The main difference with our approach is that the input probabilities have to be entered by the software architect based on his previous experience. Han [27] introduces an approach for impact analysis and change propagation based on dependencies of software artifacts. Propagation rules are defined based on change patterns. A change pattern includes initial and consequent modifications with Boolean expressions that state the dependencies of the elements involved. Han applies the approach to determine the consequent modifications in design and source code for the initial modifications in design. On the contrary, our approach supports determining impacted architectural elements for requirements changes.

Slicing techniques are mainly developed to understand dependencies using independent slices of the program [16]. Silicing is based on data and control flows in the program. Slicing techniques limit change propagation to the identification of the scope of changes. The work by Tip et al. [51] is an example of slicing techniques for C++ programs. Architectural slicing introduced by Zhao et al. [59] [60] is similar to program slicing. As opposed to program slicing, architectural slicing runs on the software architecture. The approach determines one slice of the software architecture for the proposed change. Components that might be impacted by the changed component are traced by using a graph of information flows. The slicing approach requires all the information flows of the software architecture being exposed. Zhao et al. mainly focus on the questions such as 'If a change is made to a component c, what other components might be affected by the change in c?'. On the contrary, our appraoch deals with the question 'If a change is made to a requirement r, what components might be affected by the change in r?'. Feng and Maletic [14] address the propagation of architectural changes within the same architecture. Their approach can be considered as both dependency analysis and slicing technique. Interface and method slicing are used together with analysis of component dependencies.

Westhuizen and Hoek [53] provides an approach for propagating architectural changes within a product line architecture. The approach has two algorithms. The first one is a differencing algorithm that automatically calculates the difference between two versions of a product line architecture. The second algorithm is a merging algorithm that propagates the changes, captured by the differencing algorithm, to the second product line architecture. The merging algorithm requires the presence of some common elements among the architectures. It propagates the changes from one architecture to another. On the contrary, our approach focuses on the propagation of requirements changes in one architecture.

There are multiple ADLs proposed in the literature. Each one focusses on a specific application domain, analysis type, or modelling environment, with its own notation and tools. There might be cases where the architecture is modelled with multiple ADLs. Eramo et al. [13] propose a change propagation approach between multiple architectural languages. The proposed approach ensures that when an architecture model in an ADL has been modified, such modifications are propagated in a finite number of steps to all other models in other ADLs for the same architecture. The approach [13] is complementary to our approach in cases where the architecture is modelled with multiple ADLs. The candidate architectural elements can be identified with our approach for requirements changes. After implementing changes in the architecture, the consistency of the architecture models can be ensured by the change propagation approach [13].

Byron and Carver [57] presents a systematic literature review of software architecture characteristics. The results of the literature review are used to propose a Software Architecture Characterization Scheme (SACCS). The goal of the scheme is to help the software architect to make decisions about how to address a change request. Our approach can be complemented by SACCS. After identifying candidate architectural elements with our approach, the software architect can use SACCS and candidate elements to make decisions - how to modify architecture for changes in requirements.

The approaches mentioned above focus on change impact analysis mainly within the scope of software architecture. In addition to these approaches, there are more generic impact analysis approaches which can be adapted for software architecture. Event-Based Traceability (EBT) [7] is one of these approaches which supports change impact analysis by automating trace generation and maintenance. In EBT, requirements and other traceable artifacts, such as design models, are linked through publish-subscribe relationship based on the *Observer design pattern* [17]. The main purpose of EBT is to determine candidate elements and maintain traces for these elements. Contrary to our approach, in EBT all elements directly/indirectly related to the changed element are candidate. EBT does not support identification of false positives. Cleland-Huang et al. [8] present a goal-centric approach to detect the impact of a change in non-functional requirements. Non-functional requirements and their dependencies are modeled with a Softgoal Interdependency Graph (SIG). The impact detection is limited to identifying a set of directly impacted SIG elements.

Ibrahim et al. [30] present a change impact analysis approach which contains change propagation from requirements to design, test case or source code. They consider horizontal traceability as a model of inter-artifacts such that each artifact in one abstraction level provides links to other artifacts of different abstraction levels (i.e., traces between a requirements model and software architecture). Ibrahim et al. define a requirement component relationship as a relationship between requirements and other artifacts of different levels. They, however, do not explain how to propagate a change from one requirement to other artifacts such as architecture and test cases. They only explain how to use call graphs and dependence graphs to analyze change impact between design models or between design models and source code.

Turver et al. [52] describe a technique dealing with the ripple effects of a change based on a graph-theoretic model. This technique can be applied not only for source code but also for architecture and requirements. The technique, however, calculates the ripple effects without using the semantics of traces. Chen et al. [6] introduce an impact analysis approach handling not only software contents but also other items such as requirements, documents and data. The approach traces heterogeneous items by using attributes and linkages. The linkages are between different types of items such as requirements-to-component but there is no mention of any linkage between requirements and how to use them with other linkages for change impact analysis.

In the literature there are numerous works [4] [5] [11] [12] [26] on analysis of changes for UML diagrams/models. Our approach determines the impact in software architecture when a change occurs in a requirement. Here, the requirements changes are fostered by the evolution of the business needs. In our approach, the requirements engineer mainly tries to make the architecture consistent with the business needs given in the requirements model. In the UML based approaches, the main aim is to keep the UML diagrams consistent with each other regardless of the change rationale.

## 11 Final Considerations

In this paper, we presented a technique for change impact analysis in software architecture. We use the formal semantics of requirements relations, requirements changes and traces between R&A to identify candidate architectural elements for the impact of requirements changes in the architecture. Most of the approaches and tools such as IBM Rational RequisitePro and DOORS do not focus on formal semantics of requirements relations and traces. We provide a more precise change impact analysis in software architecture which is able to rule out some false positive impacts. Designing architecture based on requirements is a creative process. There are an infinite number of designs that satisfy requirements for a given project. The number of changes over the architecture is infinite. Our approach identifies the architectural elements that implement the impacted system properties given in impacted requirements. These elements give the relevant part of the architecture for the changed/new requirements. The software architect takes candidate architectural elements as input for his/her decision about which architectural elements to be changed for the changed/new system properties. Therefore, he/she starts investigating the architecture with the candidate architectural elements in order to change it for requirements changes.

We extended our tool TRIC for our change impact analysis technique. Our tool support needs improvement for usability. The core parts of the tool are implemented. However, the integration of these parts (the integration of TRIC and the Eclipse model editor in Section 8) is currently done manually and we need a user interface to access the Eclipse model editor inside TRIC.

Our approach has limitations for some change types in some particular cases. For instance, our change impact function returns suggestion for the impact of adding a new requirement (*Add a New Requirement*) only if there is an existing requirement related to the newly added requirement. Also the function returns no suggestion for adding a new system property to an existing requirement (*Add Property to Requirement*) since there is no explicit dependency between the existing properties and the added property.

The task of identifying requirements relations during requirements modeling and generating traces between R&A is vital to our approach. In our previous work [24] [22] we thoroughly studied how to manually identify and assign the initial relations among requirements. The requirements reasoning framework given in [24] also provides a semi-automatic tool support (the reasoning features of TRIC) to infer new relations from the initial set of relations and check the consistency of the given and inferred relations. Especially, the consistency checking feature improves the correctness of the requirements relations in the model [24]. As a continuum, we presented another approach [21] [23] that provides trace establishment between R&A by using architecture verification together with the semantics of requirements relations and traces. Our approach helps the software architect to ensure the validity of traces between R&A.

# References

[1] Bohner, S. A. (2002). Extending Software Change Impact Analysis into COTS Components. *27th Annual NASA Goddard Software Engineering Workshop*, 175-182.

[2] Bohner, S. A. (2002). Software Change Impacts – An Evolving Perspective. *ICSM'02*, 263-271.

[3] Bohner, S. A., & Gracanin, D. (2003). Software Impact Analysis in a Virtual Environment. *28th Annual NASA Goddard Software Engineering Workshop*, 143-151.

[4] Briand, L. C., Labiche, Y., & O'Sullivan, L. (2003). Impact Analysis and Change Management of UML Models. *ICSM'03*, 256-265.

[5] Briand, L. C., Labiche, Y., O'Sullivan, L., & Sowka, M. (2006). Automated Impact Analysis of UML Models. *Journal of Systems and Software, 79*(3), 339-352.

[6] Chen, C. Y., & Chen, P. C. (2009). A Holistic Approach to Managing Software Change Impact. *Journal of Systems and Software, 82*(12), 2051-2067.

[7] Cleland-Huang, J., Chang, C. K., & Christensen, M. (2003). Event-based Traceability for Managing Evolutionary Change. *IEEE Transactions on Software Engineering, 29*(9), 796-810.

[8] Cleland-Huang, J., Settimi, R., BenKhadra, O., Berezhanskaya, E., & Christina, S. (2005). Goal-centric Traceability for Managing Non-functional Requirements. *ICSE'05*, 362-371.

[9] Cockburn, A. (2000). Writing Effective Use Cases. *Addison-Wesley*.

[10] Dahlstedt, A. G., & Persson, A. (2005). Requirements Interdependencies: State of the Art and Future Challenges. In A. Aurum & C. Wohlin (Eds.), *Engineering and Managing Software Requirements* (pp. 95-116). Berlin: Springer.

[11] Dam, H. K., & Winikoff, M. (2010). Supporting Change Propagation in UML Models. *ICSM'10*, 1-10.

[12] Egyed, A. (2007). Fixing Inconsistencies in UML Design Models. *ICSE'07*, 292-301.

[13] Eramo, R., Malavolta, I., Muccini, H., Pelliccione, P., & Pierantonio, A. (2012). A Model-driven Approach to Automate the Propagation of Changes among Architecture Description Languages *Software and Systems Modeling, 11*(1), 29-53.

[14] Feng, T., & Maletic, J. I. (2006). Applying Dynamic Change Impact Analysis in Component-based Architecture Design. *SNPD'06*, 43-48.

[15] Fowler, M. (1999). *Refactoring: Improving the Design of Existing Code*: Addison-Wesley.

[16] Gallagher, K. B., & Lyle, J. R. (1991). Using Program Slicing in Software Maintenance. *IEEE Transactions on Software Engineering, 17*(8), 751-761.

[17] Gamma, E., Helm, R., Johnson, R., & Vlissides, J. (1995). *Design Patterns: Elements of Reusable Object-Oriented Software*: Addison-Wesley Professional.

[18] Goknil, A. (2011). *Traceability of Requirements and Software Architecture for Change Management*. PhD Thesis, University of Twente, Enschede.

[19] Goknil, A., Kurtev, I., & Millo, J. V. (2013). A Metamodeling Approach for Reasoning on Multiple Requirements Models. *EDOC'13*, 159-166.

[20]  Goknil, A., Kurtev, I., & van den Berg, K. (2014). Change Impact Analysis in Requirements: a Metamodeling Approach. *Information and Software Technology, 56*(8), 950-972.

[21]  Goknil, A., Kurtev, I., & van den Berg, K. (2014). Generation and Validation of Traces between Requirements and Architecture based on Formal Trace Semantics. *Journal of Systems and Software, 88*, 112-137.

[22]  Goknil, A., Kurtev, I., & van den Berg, K. (2008). A Metamodeling Approach for Reasoning about Requirements. *ECMDA-FA'08,* 311-326.

[23]  Goknil, A., Kurtev, I., & van den Berg, K. (2010). Tool Support for Generation and Validation of Traces between Requirements and Architecture. *ECMFA-TW'10*, 39-46.

[24]  Goknil, A., Kurtev, I., van den Berg, K., & Veldhuis, J. W. (2011). Semantics of Trace Relations in Requirements Models for Consistency Checking and Inferencing. *Software and System Modeling, 10*(1), 31-54.

[25]  Grechanik, M., McKinley, K. S., & Perry, D. E. (2007). Recovering and Using use-case-diagram-to-source-code traceability links. *ESEC-FSE'07*, 95-104.

[26]  Groher, I., & Egyed, A. (2010). Selective and Consistent Undoing of Model Changes. *MODELS'10*, 123-137.

[27]  Han, J. (1997). Supporting Impact Analysis and Change Propagation in Software Engineering Environments. *STEP '97*, 172-182.

[28]  Hayes, J. H., Dekthyar, A., & Sundaram, S. K. (2006). Advancing Candidate Link Generation for Requirements Tracing: the Study of Methods. *IEEE Transactions on Software Engineering, 32*(1), 4-19.

[29]  Huth, M. R. A., & Ryan, M. D. (2000). Logic in Computer Science: Modeling and Reasoning about Systems. *Cambridge University Press, Cambridge.*

[30]  Ibrahim, S., Munro, M., & Deraman, A. (2005). A Requirements Traceability to Support Change Impact Analysis. *Asian Journal of Information Technology, 4*(4), 329-338.

[31]  Jonsson, P., & Lindvall, M. (2005). Impact Analysis. In A. Aurum & C. Wohlin (Eds.), *Engineering and Managing Software Requirements* (pp. 117-142). Berlin: Springer.

[32]  Khan, S. S., Greenwood, P., Garcia, A., & Rashid, A. (2008). On the Impact of Evolving Requirements-Architecture Dependencies: An Exploratory Study. *Caise'08,* 243-257.

[33]  Koch, N., & Kraus, A. (2003). Towards a common metamodel for the development of web applications. *ICWE'03*, 497-506.

[34]  Lee, M., Offutt, J. A., & Alexander, R. T. (2000). Algorithmic Analysis of the Impacts of Changes to Object-oriented Software. *TOOLS'00*, 61-70.

[35]  Li, B., Sun, X., Leung, H., & Zhang, S. (2012). A Survey of Code-based Change Impact Analysis Techniques. *Software Testing, Verification, Reliability, 23*(8), 613-646.

[36]  Lopez, O., Laguna, M. A., & Garcia, F. J. (2002). Metamodeling for requirements reuse. *WER'02*, 76-90.

[37]  Meyer, J. J. C., Wieringa, R., & Dignum, F. (1998). The Role of Deontic Logic in the Specification of Information Systems. *Logics for Databases and Information Systems*, 71-115.

[38]  Moon, M., Yeom, K., & Chae, H. S. (2005). An approach to developing domain requirements reuse as a core asset based on commonality and

variability analysis in a product line. *IEEE Transactions on Software Engineering, 31*(7), 551-569.

[39] Murta, L. G. P., van der Hoek, A., & Werner, C. M. L. (2006). ArchTrace: Policy-Based Support for Managing Evolving Architecture-to-Implementation Traceability Links. *ASE'06*, 135-144.

[40] Mylopoulos, J., Chung, L., & Yu, E. (1999). From object-oriented to goal oriented requirements analysis. *ACM Commiunications, 42*(1), 31-37.

[41] Noppen, J., van den Broek, P., & Aksit, M. (2007). Imperfect Requirements in Software Development. *REFSQ'07,* 247-261.

[42] Ölveczky, P. C., Boronat, A., & Mesequer, J. (2010). Formal Semantics and Analysis of Behavioral AADL Models in Real-Time Maude. *FMOODS/FORTE'10,* 47-62.

[43] Ölveczky, P. C., Boronat, A., Mesequer, J., & Pek, E. (to appear). Formal Semantics and Analysis of Behavioral AADL Models in Real-Time Maude. *Technical Report at UIUC*.

[44] OMG SysML Specification.   http://www.sysml.org/specs.htm

[45] Paige, R. F., Drivalos, N., Kolovos, D. S., Fernandes, K. J., Power, C., Olsen, G. K., et al. (2011). Rigorous Identification and Encoding of Trace-links in Model-Driven Engineering. *Software and Systems Modeling, 10*(4), 469-487.

[46] Ramesh, B., & Jarke, M. (2001). Towards reference Models for Requirements Traceability. *IEEE Transactions on Software Engineering, 27*(1), 58-93.

[47] Rashid, A., Moreira, A., & Araujo, J. (2003). Modularization and composition of aspectual requirements. *AOSD'03*, 11-20.

[48] Architecture Analysis and Design Language (AADL). http://www.aadl.info

[49] SWEBOOK: Guide to Software Engineering Body of Knowledge. *IEEE Computer Society*.

[50] Tang, A., Jin, Y., Han, J., & Nicholson, A. (2005). Predicting Change Impact in Architecture Design with Bayesian Belief Networks. *WICSA'05*, 67-76.

[51] Tip, F., Jong, D. C., Field, J., & Ramlingam, G. (1996). Slicing Class Hierarchies in C++. *OOPSLA'96*, 179-197.

[52] Turver, R. J., & Munro, M. (1994). An Early Impact Analysis Technique for Software Maintenance. *Journal of Software Maintenance Research and Practice, 6*(1), 35-52.

[53] van der Westhuizen, C., & van der Hoek, A. (2002). Understanding and Propagating Architectural Changes. *WICSA'02*, 95-109.

[54] van Lamswerdee, A. (2001). Goal-oriented requirements engineering: a roundtrip from research to Practice. *RE'01*, 249-263.

[55] Veldhuis, J. W. (2009). *Tool support for a metamodeling approach for reasoning about requirements.* MSc Thesis, University of Twente, Enschede.

[56] Wasson, C. S. (2006). *System, Analysis, Design, and Development: Concepts, Principles, and Practices*: John Wiley & Sons.

[57] Williams, B. J., & Carver, J. C. (2010). Characterizing Software Architecture Changes: A Systematic Review. *Information and Software Technology, 52*(1), 31-51.

[58] Zhang, H., Li, J., Zhu, L., Jeffery, R., Liu, Y., Wang, Q., et al. (2014). Investigating Dependencies in Software Requirements for Change

Propagation Analysis. *Information and Software Technology, 56*(1), 40-53.

[59] Zhao, J. (1998). Applying Slicing Technique to Software Architectures. *ICECCS'98*, 87-98.

[60] Zhao, J., Yang, H., Xiang, L., & Xu, B. (2002). Change Impact Analysis to Support Architectural Evolution. *Journal of Software Maintenance: Research and Practice, 14*(5), 317-333.

# Appendix 1. Abbreviations of the Elements in the RPM System

| Abbreviation | Explanation |
|---|---|
| SD | Sensor Device |
| SDC | Sensor Device Coordinator |
| SDM | Sensor Device Manager |
| AS | Alarm Service |
| AR | Alarm Receiver |
| WS | Web Server |
| WC | Web Client |
| HPC | Host Personal Computer |
| CPC | Client Personal Computer |
| sd_blood_edp1 | Event Data Port 1 for Blood Pressure in Sensor Device |
| sd_blood_edp2 | Event Data Port 2 for Blood Pressure in Sensor Device |
| sd_blood_edp3 | Event Data Port 3 for Blood Pressure in Sensor Device |
| sd_blood_edp4 | Event Data Port 4 for Blood Pressure in Sensor Device |
| sd_temp_edp1 | Event Data Port 1 for Temperature in Sensor Device |
| sd_temp_edp2 | Event Data Port 2 for Temperature in Sensor Device |
| sd_temp_edp3 | Event Data Port 3 for Temperature in Sensor Device |
| sd_temp_edp4 | Event Data Port 4 for Temperature in Sensor Device |
| sd_temp_alarm_edp1 | Event Data Port 1 for Temperature Alarm in Sensor Device |
| sd_temp_alarm_edp1 | Event Data Port 1 for Temperature Alarm in Sensor Device |
| sd_temp_alarm_edp3 | Event Data Port 3 for Temperature Alarm in Sensor Device |
| sd_temp_alarm_edp4 | Event Data Port 4 for Temperature Alarm in Sensor Device |
| sdThr | Thread in Sensor Device |
| sdc_blood_edp1 | Event Data Port 1 for Blood Pressure in Sensor Device Controller |
| sdc_blood_edp2 | Event Data Port 2 for Blood Pressure in Sensor Device Controller |
| sdc_blood_edp3 | Event Data Port 3 for Blood Pressure in Sensor Device Controller |
| sdc_blood_edp4 | Event Data Port 4 for Blood Pressure in Sensor Device Controller |
| sdc_blood_edp5 | Event Data Port 5 for Blood Pressure in Sensor Device Controller |
| sdc_blood_edp6 | Event Data Port 6 for Blood Pressure in Sensor Device Controller |
| sdc_temp_edp1 | Event Data Port 1 for Temperature in Sensor Device Controller |
| sdc_temp_edp2 | Event Data Port 2 for Temperature in Sensor Device Controller |
| sdc_temp_edp3 | Event Data Port 3 for Temperature in Sensor Device Controller |
| sdc_temp_edp4 | Event Data Port 4 for Temperature in Sensor Device Controller |
| sdc_temp_edp5 | Event Data Port 5 for Temperature in Sensor Device Controller |
| sdc_temp_edp6 | Event Data Port 6 for Temperature in Sensor Device Controller |
| sdc_temp_alarm_edp1 | Event Data Port 1 for Temperature Alarm in Sensor Device Controller |
| sdc_temp_alarm_edp2 | Event Data Port 2 for Temperature Alarm in Sensor Device Controller |
| sdc_temp_alarm_edp3 | Event Data Port 3 for Temperature Alarm in Sensor Device Controller |
| sdc_temp_alarm_edp4 | Event Data Port 4 for Temperature Alarm in Sensor Device Controller |
| sdc_temp_alarm_edp5 | Event Data Port 5 for Temperature Alarm in Sensor Device Controller |
| sdc_temp_alarm_edp6 | Event Data Port 6 for Temperature Alarm in Sensor Device Controller |
| sdcThr | Thread in Sensor Device Controller |
| sdm_blood_edp1 | Event Data Port 1 for Blood Pressure in Sensor Device Manager |
| sdm_blood_edp2 | Event Data Port 2 for Blood Pressure in Sensor Device Manager |
| sdm_blood_strg | Storage for Blood Pressure in Sensor Device Manager |
| sdm_temp_edp1 | Event Data Port 1 for Temperature in Sensor Device Manager |
| sdm_temp_edp2 | Event Data Port 2 for Temperature in Sensor Device Manager |
| sdm_temp_strg | Storage for Temperature in Sensor Device Manager |
| sdm_temp_alarm_edp1 | Event Data Port 1 for Temperature Alarm in Sensor Device Manager |
| sdm_temp_alarm_edp2 | Event Data Port 2 for Temperature Alarm in Sensor Device Manager |
| sdm_temp_alarm_strg | Storage for Temperature Alarm in Sensor Device Manager |
| sdmThr | Thread in Sensor Device Manager |
| hpc_blood_edp1 | Event Data Port 1 for Blood Pressure in Host Personal Computer |
| hpc_temp_edp1 | Event Data Port 1 for Temperature in Host Personal Computer |
| hpc_temp_req_edp1 | Event Data Port 1 for Temperature Request in Host Personal Computer |
| hpc_temp_alarm_edp1 | Event Data Port 1 for Temperature Alarm in Host Personal Computer |
| wc_temp_req_edp1 | Event Data Port 1 for Temperature Request in Web Client |
| wc_temp_req_edp2 | Event Data Port 2 for Temperature Request in Web Client |
| wc_temp_req_edp3 | Event Data Port 3 for Temperature Request in Web Client |
| wc_temp_req_edp4 | Event Data Port 4 for Temperature Request in Web Client |
| wcThr | Thread in Web Client |
| ws_temp_req_edp1 | Event Data Port 1 for Temperature Request in Web Server |
| ws_temp_req_edp2 | Event Data Port 2 for Temperature Request in Web Server |
| ws_temp_req_edp3 | Event Data Port 3 for Temperature Request in Web Server |
| ws_temp_req_edp4 | Event Data Port 4 for Temperature Request in Web Server |
| wsThr | Thread in Web Server |
| cpc_temp_req_edp1 | Event Data Port 1 for Temperature Request in Client Personal Computer |
| cpc_temp_req_edp2 | Event Data Port 2 for Temperature Request in Client Personal Computer |
| cpc_ar | Alarm receiver in Client Personal Computer |

# Appendix 2. Graphical Notation of AADL

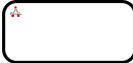 System      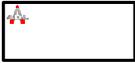 Datastore

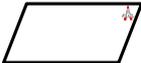 Process      Event Data Port

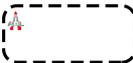 Thread Group   Data Access

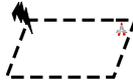 Thread       Connector

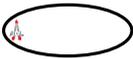 Subprogram